\documentclass[%
unsortedaddress,
twocolumn,
amsmath,amssymb,
aps,
prb]{revtex4-2}
\usepackage{siunitx}
\usepackage[italicdiff]{physics}
\usepackage{mathtools}
\usepackage{wrapfig}
\usepackage{xurl}
\usepackage{array}
\usepackage{amsmath}
\usepackage{amsthm}
\usepackage{graphicx}
\usepackage{dcolumn}
\usepackage{bm}
\usepackage{color}
\usepackage[normalem]{ulem}
\usepackage[makeroom]{cancel}
\newcommand{\iu}{{\rm i}}

\begin{document}
\title{Spin parity effects in monoaxial chiral ferromagnetic  chain}

\author{Sohei Kodama}
 \email{kodama.s.0411@gmail.com}
 \affiliation{Department of Basic Science, The 
University of Tokyo, Komaba 3-8-1, Meguro, Tokyo 153-8902, Japan}
\author{Akihiro Tanaka}
\email{TANAKA.Akihiro@nims.go.jp}
\affiliation{International Center for Materials Nanoarchitectonics, National Institute for Materials Science, 1-1 Namiki, Tsukuba, Ibaraki 305-0044, Japan}
\author{Yusuke~Kato}
 \email{yusuke@phys.c.u-tokyo.ac.jp}
\affiliation{Department of Basic Science, The 
University of Tokyo,  Komaba 3-8-1, Meguro,Tokyo 153-8902, Japan}
%
\date{\today}
\begin{abstract}
While spin parity effects– physics 
crucially depending on whether the spin quantum number $S$ is half-odd integral or integral, 
 have for decades been a source of new developments for the quantum physics of antiferromagnetic spin chains, the investigation into their possible ferromagnetic counterparts have remained largely unchartered, especially in the fully quantum (as opposed to the semiclassical) regime. Here we present such studies for 
 monoaxial chiral ferromagnetic spin chains.
We start by examining 
magnetization curves for finite-sized systems,  
where a magnetic field is applied perpendicular to the helical axis. 
For half-odd integer $S$, the curves feature discontinuous jumps  
identified as a series of level crossings, each accompanied by a shift of the crystal momentum $k$ by an amount of $\pi$.
The corresponding curves for integer-valued $S$ are continuous  
and exhibit crossover processes. For the latter case $k=0$ throughout. 
These characteristics 
are observed numerically when the strength of the 
Dzyaloshinskii-Moriya interaction (DMI) $D$ is comparable to or larger than that of the 
ferromagnetic exchange interaction $J$. 
Solitons are known to be responsible for step-wise changes seen in magnetization curves in the classical limit. These findings therefore prompt us to 
revise the notion of a soliton, for arbitrary $S$, into a quantum mechanical entity. 

To unravel this phenomenon at the fully quantum 
level as is appropriate to spin chains with small $S$, 
 we examine in detail special limiting Hamiltonians amenable to rigorous analysis,  consisting of only the DMI and the Zeeman energy. Dubbed the $DH$ model (for $S=1/2$) and the projected $DH$ (p$DH$) model (for general $S$), they have a set of $2S$ conserved quantities, each of which is the number of solitons of a specific integer-valued height (as measured in the $S_z$ basis), which ranges from 1 to $2S$. We discuss how to determine the exact crystal momentum of the lowest energy state belonging to a sector with a given set of the $2S$ soliton numbers.   Combined with energetic considerations, this information enables us to reproduce the spin parity effect in the magnetization curves. Finally we show that the ground state of the special models have substantial numerical overlap with those for generic systems with a finite exchange interaction, suggesting the same physics to be valid there as well. 
\end{abstract}
\maketitle
\section{Introduction}
It has long been known that an asymmetric exchange interaction, the Dzyaloshinskii-Moriya interaction (DMI)\cite{dzyaloshinskyThermodynamicTheoryWeak1958a,moriyaAnisotropicSuperexchangeInteraction1960a}, 
is allowed for pairs of spins bridged with bonds lacking an inversion center. 
Within a classical treatment of spins, this interaction acts in such a way as to twist the relative orientation of the adjacent spin moments. 
Owing to this feature, a competition between symmetric exchange interactions  
and DMI can induce 
topologically nontrivial configurations such as skyrmions\cite{bogdanov1989,
bogdanovThermodynamicallyStableMagnetic1994,Papas2009,mb2009skyrmion,yu2010,nagaosa2013topological} and chiral solitons\cite{dzyaloshinskii1965theory,moriyamiyadai1982,miyadaiMagneticPropertiesCr1983,zheludev1997field,togawa2012,kishineovchinnikov2015,togawa2016}, 
which will largely behave as stable, particle-like entities with a fixed chirality. That these emergent particles are very much real was demonstrated forcefully in the past decade through 
 Lorentz transmission spectroscopy experiments\cite{yu2010,togawa2012}, after which an extensive exploration into their thermodynamics and transport properties ensued. 

 There exist in the literature studies which highlight the significance of the effect of the DMI on {\it quantum} spin systems, 
e.g. in coupled spin chains modeling CsCuCl$_3$\cite{NikuniShiba1993} and in $S=1/2$ antiferromagnetic spin chains modeling Cu-benzoate \cite{OshikawaAffleck1997,AffleckOshikawa1999}. It is also true though that the DMI's role  
becomes somewhat elusive once one opts for a fully quantum mechanical treatment of the magnet ---an essential requirement when $S$, the spin quantum number,  
is small. This owes to the fact that the notion of a classical spin vector breaks down in this limit, 
implying that the intuitive understanding deriving from a (semi-)classical picture  (valid for sufficiently large $S$) 
is no longer at our disposal.  

Some time ago, Braun and Loss performed their pioneering study on the quantum dynamics of solitons in 
effectively one dimensional nanomagnets in the {\it absence} of a DMI\cite{BraunLoss1996}. 
Though emerging out of a nonchiral magnet, chirality turns out to play 
an essential role for solitons which are stabilized in such systems 
by anisotropic interactions. Using semiclassical methods 
while taking the crucial step of keeping track of spin Berry phases, 
it was shown how the latter 
gives rise to {\it spin parity effects} 
-- the dependence of the system's behavior on the parity of twice the spin quantum number 2$S$:  
the Bloch bands formed by solitons 
exhibit different structures for integer and half-odd integer $S$, and as a consequence, tunneling between opposite chiralities can occur for the half-odd integer $S$ case. 
These authors went on to a provide a separate analysis\cite{BraunLossIntJModPhysB1996} for $S=1/2$ quantum spin chain models (for both ferromagnets and antiferromagnets) with the same symmetry, where within Villain's approximation\cite{villain1975} 
of projecting to a fixed soliton number sector, they found results   
that are consistent with their semiclassical treatment. 

Turning to {\it chiral} magnets, a similar 
semiclassical investigation was undertaken by 
Takashima {\it et al}\cite{Takashima2016} in their work on skyrmion dynamics in 2D chiral  ferromagnets, 
also pointing to a spin parity effect. 
The major conclusion drawn was that the lowest energy state 
in the sector with $N_{\rm s}$ skyrmions acquires a crystal momentum of ${\bm k}=(2\pi S N_{\rm s}, -2\pi S N_{\rm s})$. 
i.e., ${\bm k}$ resides at the zone edge when $S N_{\rm s}$ is half-integral, while 
being located  at the zone center when $S N_{\rm s}$ is integer-valued. 
It was further argued that the same dichotomy manifests itself when one examines the phase diagram of the system as a function of the applied magnetic field. 

Historically, spin parity effects related to topologically nontrivial configurations in quantum spin systems came into focus with the work of Haldane\cite{HaldanePLA1983,HaldanePRL1988} on the spectral properties of antiferromagnetic spin chains. (It is worth mentioning that relations to the crystal momentum of soliton states\cite{HaldanePLA1983} and hedgehog processes\cite{HaldanePRL1988}  were also briefly addressed in the course of these studies.) 
The aforementioned body of work suggests 
that they can arise in a wider range of systems, often with intriguing implications. 

With the sole exception of antiferromagnetic spin chains which have been thoroughly scrutinized, it is still largely unknown though, what the full structure of the {\it quantum limit} theory and its implications are for most of the problems mentioned above. 
The present work was motivated largely by 
the recent advent of the physics of  
monoaxial chiral ferromagnetic chains\cite{kishineovchinnikov2015, togawa2016}. 
As their analysis to date had mainly been conducted within the (semi-)classical micromagnetic framework, we view the undertaking of its study from a {\it purely quantum} perspective  
an important and urgent task. This explains the purpose of the present work.  

In the following sections we will be dealing with quantum spin chains models 
of chiral ferromagnets for arbitrary $S$. We begin by examining the numerically obtained magnetization curves for the cases  
$S=1/2, 1, 3/2$ and $2$.  
Previously the magnetization curves of finite-sized monoaxial spin chains were calculated in 
\cite{kishineTopologicalMagnetizationJumps2014} in the {\it classical} case, where step-wise changes appearing in the curves were ascribed to solitons.  
Our numerical results for {\it 
 quantum} spin chains show a new feature: a prominent spin parity effect is at play, as detailed in later sections. With the semiclassical picture for solitons unavailable, however, it is not immediately apparent why this should be so. Nor is it obvious what exactly the notion of a soliton becomes when treated as a quantum mechanical entity. 
We will show that much insight into these problems   
is gained through the {\it exact} study of a limiting case of our full Hamiltonian which we dub the $DH$ model, wherein only the DMI and the Zeeman energy are retained. A variant which we will call the projected $DH$ (p$DH$) model will prove to be invaluable when we turn to the $S>1/2$ cases. 
These models allow for a clear and rigorous understanding of how the observed spin parity effect 
can be reproduced 
in terms of solitons that are well-defined in the quantum limit. 
While a situation in which the DMI far exceeds the symmetric exchange interaction is admittedly artificial, we will provide numerical evidence strongly suggesting that 
the $DH$ and p$DH$ models nevertheless capture a generic feature of chiral spin chains. 

Perhaps the best way to recap the foregoing paragraph is to view the $DH$ and p$DH$ models as 
 {\it parent Hamiltonians}, which generate canonical quantum-soliton states whose exact properties can be utilized to understand the physics of a whole family of generic quantum spin chains. 
 Here we are able to see a parallel structure with the Haldane gap phenomenon in antiferromagnetic spin chains, where the exactly solvable AKLT model\cite{AKLT_PRL_1987,AKLT_CMP_1988}, acting as the  parent Hamiltonian of valence-bond-solid states, played an instrumental role in our understanding of this important spin parity effect at the fully quantum level. 
 In the appendix, we provide a semiclassical account of our problem. While this 
 approach is justified in a parameter regime which is distinct from that of our quantum limit theory, 
 notable similarities in the arguments and outcomes can be detected. 
 This again is reminiscent of the situation for the antiferromagnetic counterpart, where the dual viewpoints deriving from the semiclassical theory of Haldane taken together with the AKLT picture, helped to establish a firmer understanding of the subject. It is worth noting in particular that the quantum picture owing to AKLT has since proved to be essential in bringing this understanding to new heights, where the AKLT state was shown to be a prototype of symmetry-protected-topological states\cite{Pollmann_SPT_PRB_2012}, as well as 
 a canonical platform for performing a measurement-based 
 quantum computation\cite{Miyake_AKLT_PRL_2010}. 
 We think that similar developments may well be in store for chiral ferromagnets and their quantum solitons.

We list below our main findings:
\begin{itemize}
    \item In the $DH$ model for $S=1/2$, 
    the number of solitons $(=N) $ is a good quantum number. 
A single soliton has a crystal momentum  $\pi$ at its minimum energy state. 
The lowest energy state of the $DH$ model within the sector of states containing $N$ solitons has the crystal momentum $\pi N$.   
The need to study the general $S$ case naturally lead us to introduce a variant of the $DH$ model (p$DH$ model), in which 
  the numbers of solitons $(=N_f)$ of  height $f=1,2,\cdots 2S$ are all good quantum numbers. 
  Each soliton 
  of height $f$ has the crystal momentum $\pi f$ at its minimum energy state, viz., the existence/absence of a  $\pi$-shift in the crystal momentum depends on the {\it parity of the height} of each soliton. 
  
  \item Numerical calculations imply that solely the solitons with maximal height $f=2S$ contribute to 
  the ground states in the p$DH$ model for general $S$. 
  Such energetics, together with the height parity effect conspire to 
  cause 
  the spin parity effect, $k=2\pi S N_{2S}$ in the ground state of the p$DH$ model. 
    \item For $S=1 (S=3/2)$, we find a $0.97$ ($0.91$) overlap between the ground state of the $DH$ model and p$DH$ model 
    throughout the relevant range of the magnetic field. 
    \item For $S=1/2 (S=1)$, 
    we find that probability of states with one-soliton with height 1 (2) in the ground state for the chiral magnet with $J=D$ is $80\%$ ($58\%$) slightly below the critical field.   
\end{itemize}

The remainder of the paper is organized as follows:
In the next section, we explain our model for the monoaxial chiral magnet. 
Section~\ref{sec: numerics} 
discusses numerical results on the magnetization process of finite-sized systems. They will be used 
to set the issues to be addressed in later sections. 
Section~\ref{sec: s=1/2} focuses on the study of $S=1/2$ chiral magnets in the limit $J\rightarrow 0$. 
We turn to higher $S$ cases in Section~\ref{sec: higher s}. 
Sections~\ref{sec: chirality} and \ref{sec: J} are devoted, respectively, to the chirality of quantum solitons and the influence of exchange interactions. 
Taking stock of what we have learned, in Section~\ref{sec: discussion}, we discuss the implications of the results obtained, offering an intuitive picture for them and pointing to future problems.
We state our conclusions in section~\ref{sec: conclusion}. 
The appendix  
discusses how the   
semiclassical approach applies to the 1d monoaxial chiral magnet. 
To streamline our discussion, some of the more technical matter are 
relegated to the Supplemental Material\cite{sm}, where readers will find the proofs of various lemmas stated in Sections~\ref{sec: s=1/2} and \ref{sec: higher s}, and well as detailed calculations that 
verify several of the results of these sections.

\section{Model}
In this paper we will be concerned with the ground states of quantum spin chain models of monoaxial chiral ferromagnets. Our Hamiltonian in its most complete form reads
\begin{align}
&\hat{\mathcal{H}}_{\rm ch}\nonumber\\
&=\sum_i\left[-J \hat{\bm{S}}_i \cdot\hat{\bm{S}}_{i+1}+D\left(\hat{\bm{S}}_i \times\hat{\bm{S}}_{i+1}\right)^y -H\hat{S}_i^z+K\left(\hat{S}_i^y\right)^2\right],
\label{eq: Hch}
\end{align}
where the $J$-, $D$-, and $H$-terms each stand for the exchange, the Dzyaloshinskii-Moriya and the Zeeman interactions, and the $K$-term is the single ion anisotropy. The chiral axis 
($y$-axis) is chosen to coincide with the extent of the spin chain, while the magnetic field is applied perpendicular to it. 
We set $J$, $H$, $D$, $K$ to be non-negative. 
Much of the discussions to follow will 
be devoted to the study of the limit $J=K=0$ in Eq.~\eqref{eq: Hch}. 
We will refer to the corresponding Hamiltonian as the $DH$ model:
\begin{align}
\hat{\mathcal{H}}_{DH}=
\hat{\mathcal{H}}_{\rm DM}+\hat{\mathcal{H}}_{\rm Z},
\label{eq: HDH}
\end{align}
in which
\begin{align}
&\hat{\mathcal{H}}_{\rm DM}=D\sum_j \left(\hat{\bm{S}}_j \times\hat{\bm{S}}_{j+1}\right)^y\label{eq: HDM}\\
&\hat{\mathcal{H}}_{\rm Z}=-H\sum_j\hat{S}_j^z.\label{eq: HZ}
\end{align}
We choose the number of sites $L$ to be even and impose periodic boundary conditions throughout this paper. 
The wavefunction of a finite-sized spin chain with $L$ sites  
will be expressed in terms of the  
ortho-normal basis 
$|n_1,n_2,\cdots,n_L\rangle=:|{\bm n}\rangle$, where the entries 
$n_i=0,1,\cdots 2S$ ( $i=1,\cdots, L$) are defined through the relation
$\hat{S}_{i,z}|\bm{n}\rangle=(S-n_i)|\bm{n}\rangle$. 
(We caution the reader that this is therefore {\it not} the usual $S_z$ basis.) 
The fully polarized state under a large magnetic field (commonly referred to as the forced ferromagnetic state), for example, is represented by the vector $|00\cdots00\rangle$. 
Below, for the sake of clarity, we will often exemplify general discussions on basis vectors in terms of specific spin configurations. 
The site-translation operator $\hat{T}$ acts on this basis as
$\hat{T}|\bm{n}\rangle=|T(\bm{n})\rangle$ where $T(\bm{n}):=(n_2,n_3,\cdots,n_L,n_1)$.
We denote multiple actions of $T$ on $\bm{n}$ by $T^{l}(\bm{n})=T(T^{l-1} (\bm{n}))$ for positive integer $l$.  
Consider, as a specific example of a basis vector in this representation, the state   
\begin{equation}
|00121100012210\rangle.
\label{eq: motif}
\end{equation}
When acted on by $\hat{T}$, this transforms as
\begin{equation}
\hat{T}|00121100012210\rangle=|00012110001221\rangle.
\label{eq: action of T}
\end{equation}
A word on conventions for site indices: appearances of $n_i$ for $i\notin [1,L]$ below are understood to mean 
$n_{i'}$ for $i' \in [1,L]$ such that $|i-i'|\equiv 0$ (mod $L$). 
For example, $n_0, n_{-1}$ and $n_{L+1}$ should each read as $n_L$,  $n_{L-1}$, and $n_{1}$.
%

\section{Numerical results for finite sized systems}\label{sec: numerics}

\begin{figure}
\includegraphics[width=0.9\columnwidth,pagebox=artbox]{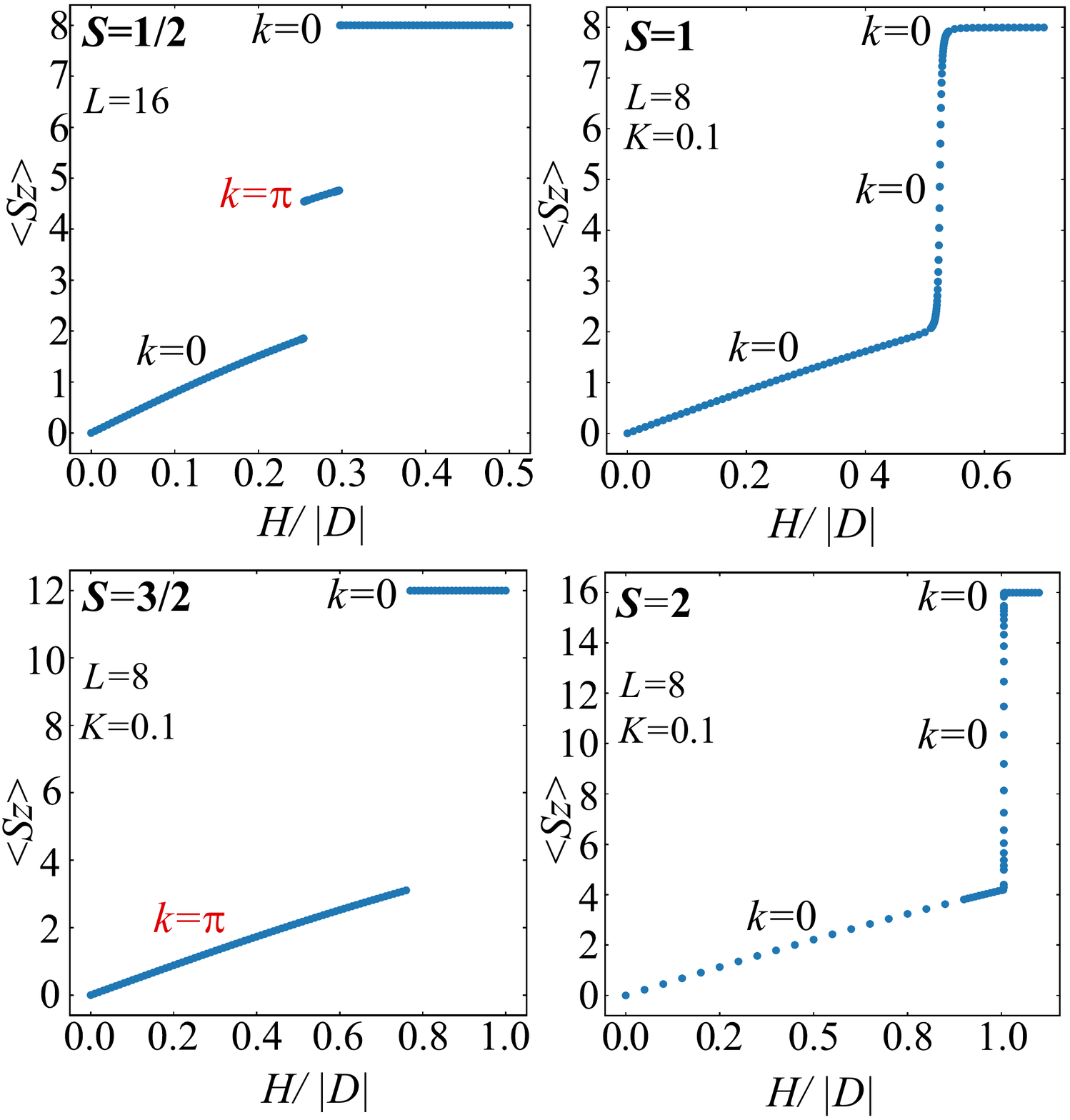}

\caption{
Magnetization curves of finite-sized spin chains for  
$S=1/2,1,3/2,$ and 2, where we set $D/J=1$.}
\label{fig: MH_D1}
\end{figure}

\begin{figure}
\includegraphics[width=0.9\columnwidth,pagebox=artbox]{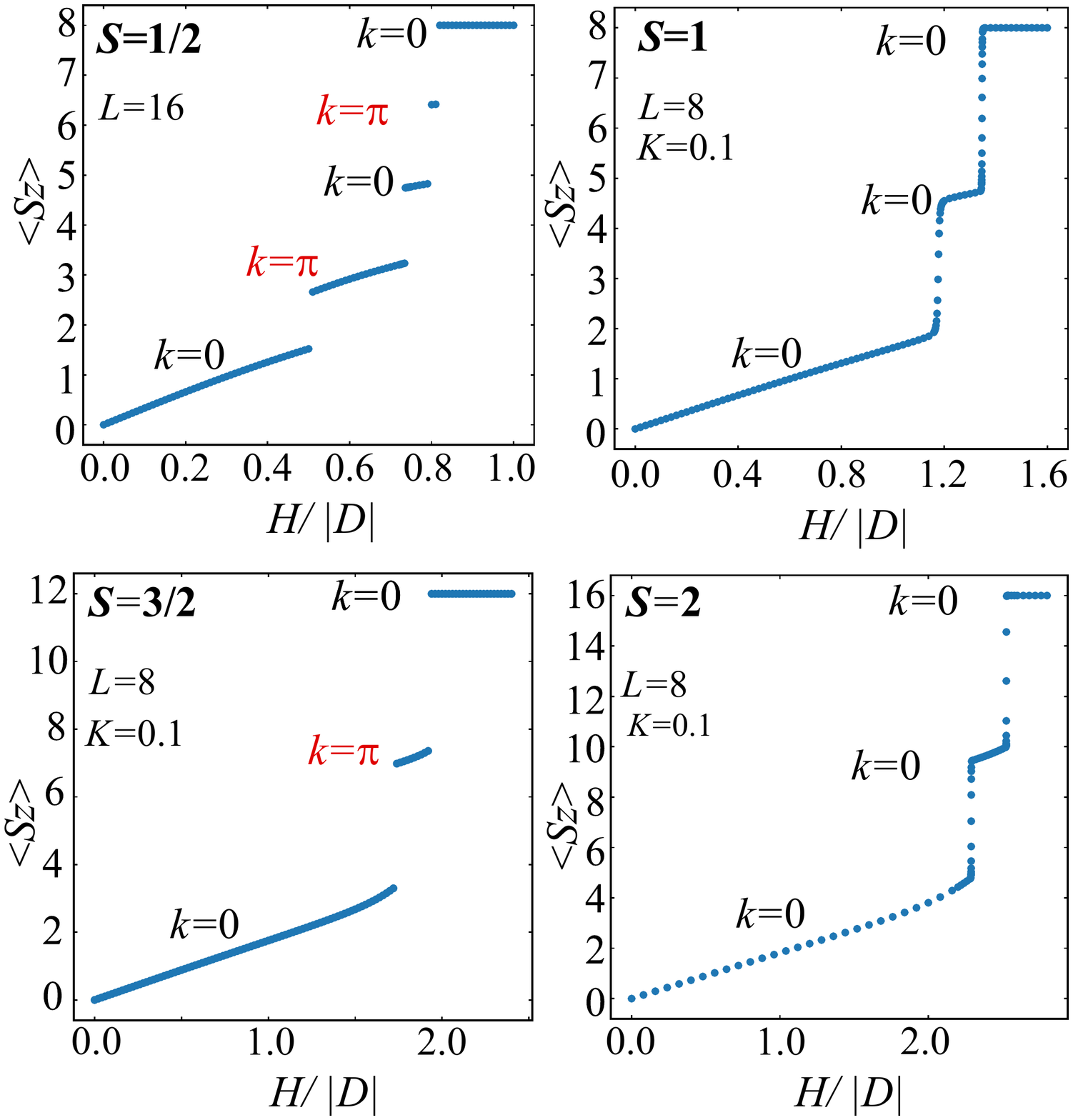}
\caption{\label{fig: MH_D50} 
Magnetization curves of finite-sized spin chains 
for $S=1/2,1,3/2,$ and 2. Here we have set $D/J=50$.}
\end{figure}

Figures~\ref{fig: MH_D1} and \ref{fig: MH_D50} are 
numerical results for the magnetization curve of finite-sized systems for various spin quantum numbers ($S=1/2,1,3/2,2$). They were obtained through the exact diagonalization of $\hat{\mathcal{H}}_{\rm ch}$. For each $S$ the curves are displayed for the two cases, $D=J$ and $D=50J$. 
As noted earlier the magnetization curves of finite-sized {\it classical} monoaxial spin chains were calculated in \cite{kishineTopologicalMagnetizationJumps2014}. 
The magnetization for $S=1/2$ and $S=3/2$ exhibit discontinuities 
as a function of the magnetic field. As indicated in the figures, they 
correspond to level crossings accompanied by a $\pi$ shift of the 
crystal momentum. The data for $S=1$ and $S=2$ show a strikingly different behavior. 
They are continuous and exhibit one or several crossovers. The crystal momentum 
of the ground state continues to be zero throughout the entire curve.

Motivated by the observation that the features mentioned above persist irrespective of the ratio $D/J$  
so long as $D/J\ge 1$, we focus in the following 
two sections on the limiting case with $J=0$ and finite $D$ (i.e. the $DH$ model (\ref{eq: HDH}) and 
its natural extension to arbitrary $S$, the projected $DH$ (or p$DH$, for short) model, which will appear in Eq.(\ref{eq: H_M}).) 
Remarkably, it turns out that with these models,  
the mechanism underlying the different behaviors between half-odd integer $S$ and integer $S$  becomes completely tractable. The effect of a finite $J$ is discussed in Sec.~\ref{sec: J}, following an exact analysis of the $DH$  and p$DH$ models.

\section{$S=1/2$ model in the limit $J\rightarrow 0$ with finite $D$  \label{sec: s=1/2}.}
\subsubsection{Basis and Conserved Quantities}
When a periodic boundary condition is imposed on 
the $DH$ model Eq.~\eqref{eq: HDH} for the case $S=1/2$, 
one can show that the eigenvalue of the operator 
\begin{equation}
\hat{N}=\sum_{i=1}^L \left(\frac14-\hat{S}_i^z\hat{S}_{i+1}^z\right)\label{eq: N-for-S=1/2}
\end{equation}
is a conserved quantity. 
One also sees by experimenting with specific examples that this operator counts one half the number of pairs of antiparallel spins 
occupying adjacent sites. For example, 
\begin{align}
&\hat{N}|0001111100\rangle= |0001111100\rangle,\label{eq: N=1example}
\\
&\hat{N}|0011100110\rangle=2|0011100110\rangle.\label{eq: N=2example}
\end{align}
In this section we shall call consecutive entries of $^^ ^^ 1"$ in the sea of $^^ ^^ 0"$'s a {\it soliton}. (Extensions of this notion 
to higher $S$ cases will be the subject of later sections.) 
The action of $\hat{N}$ on states can then be regarded as the counting of the soliton number. We denote the set of $\bm{n}$'s such that $\hat{N}|\bm{n}\rangle=N|\bm{n}\rangle$ by $V_N$. 

To see that Eq.~\eqref{eq: N-for-S=1/2} is 
conserved as announced, we note that 
the Zeeman energy and $\hat{N}$ can  
both be 
expressed in terms of $\{\hat{S}_{i,z}\}$ only. 
The two quantities therefore commute. 
The commutativity between $\hat{N}$ and the remaining DMI  can also be checked  
by direct calculation as we now show. 
To this end, as well as for later discussions, 
it proves convenient to rewrite the DMI 
in the following way: 
\begin{equation}
\hat{\mathcal{H}}_{\rm DM}=-D\sum_{i=1}^L \hat{h}_i,\label{eq: HDM0-1/2}    
\end{equation}    
where
\begin{equation}
\hat{h}_i=\frac12\left(\hat{S}_{i+1,z}-\hat{S}_{i-1,z}\right)
\left(\hat{S}_{i,+}+\hat{S}_{i,-}\right),\label{eq: hi}
\end{equation}    
with $\hat{S}_{i,\pm}=\hat{S}_{i,x}\pm i \hat{S}_{i,y}$. 
Once written in this form, it becomes clear that 
acting on a state $|\bm{n}\rangle$ with 
$\hat{h}_i$ 
will result in a null vector unless  
$n_{i+1}\ne n_{i-1}$. 
Situations where this non-vanishing condition is met 
are exhausted by the following four cases:
\begin{subequations}
\begin{align}
&2\hat{h}_i |\cdots0\overset{i}{0}1\cdots \rangle=-|\cdots0\overset{i}{1}1\cdots \rangle\label{eq: action-of-hi-a}\\    
&2\hat{h}_i |\cdots0\overset{i}{1}1\cdots \rangle=-|\cdots0\overset{i}{0}1\cdots \rangle\label{eq: action-of-hi-b}\\    
&2\hat{h}_i |\cdots1\overset{i}{0}0\cdots \rangle=+|\cdots1\overset{i}{1}0\cdots \rangle\label{eq: action-of-hi-c}\\    
&2\hat{h}_i |\cdots1\overset{i}{1}0\cdots \rangle=+|\cdots1\overset{i}{0}0\cdots \rangle\label{eq: action-of-hi-d}.    
\end{align}
\end{subequations}
A close inspection of Eqs.~\eqref{eq: action-of-hi-a}-\eqref{eq: action-of-hi-d}, reveals that $\hat{h}_i$ generates a nonzero state vector 
when site $i$ is located at the boundary between a soliton $^^ ^^ 1111"$ and the background $^^ ^^ 0000"$, in which case the new state 
has a boundary that has been shifted by one site to the left (\eqref{eq: action-of-hi-a},\eqref{eq: action-of-hi-d}) or to the right (\eqref{eq: action-of-hi-b},\eqref{eq: action-of-hi-c}).
In all four cases the length of the soliton has changed while 
the number of solitons is preserved. 
The sum of $\hat{h}_i$, Eq.~\eqref{eq: HDM0-1/2} therefore commutes with $\hat{N}$ under a periodic boundary condition.   
\subsubsection{Finite size calculation of Magnetization and Spectrum}
\begin{figure}
\includegraphics[width=\columnwidth,pagebox=artbox]{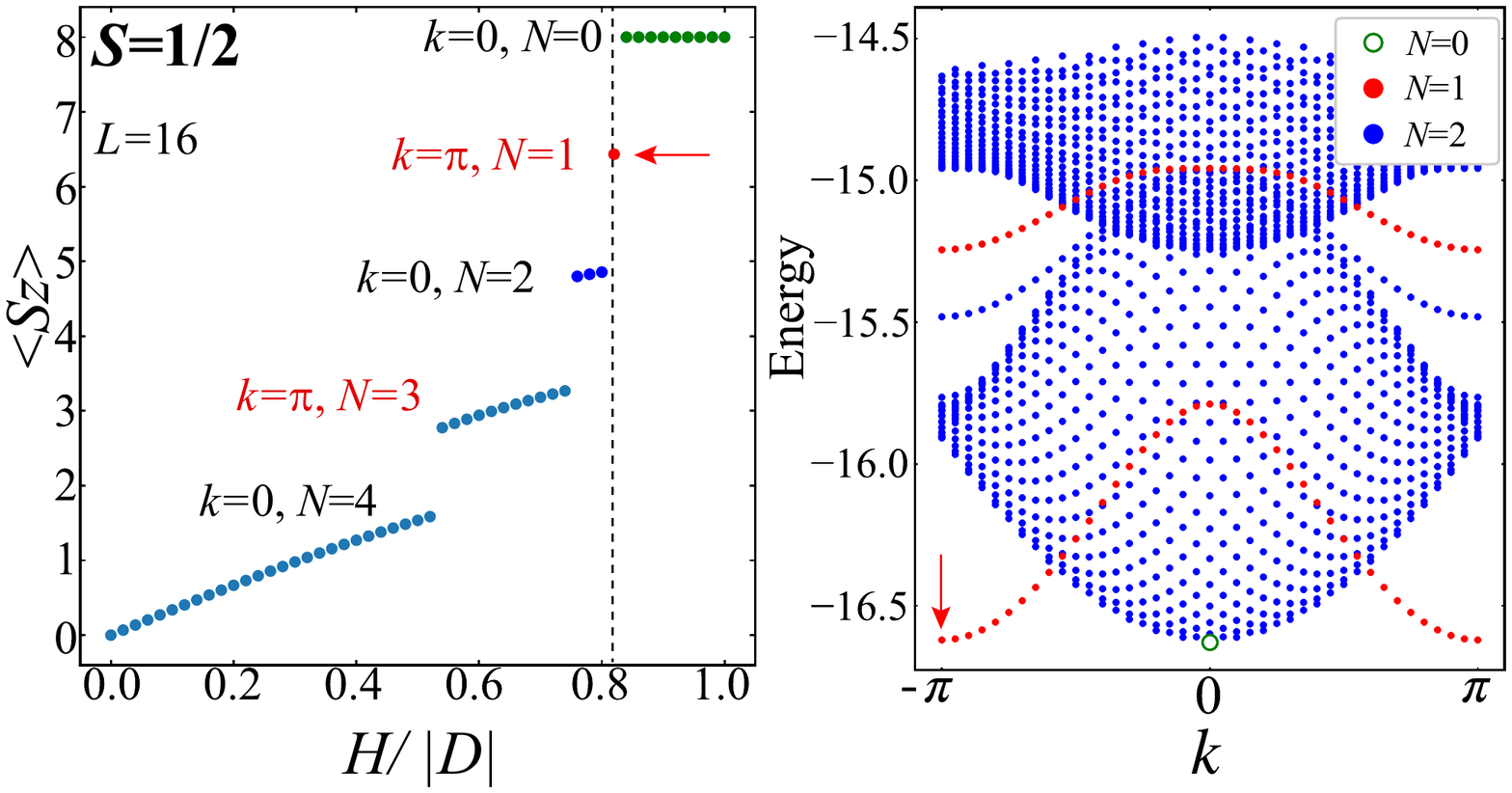}
\caption{ 
(Left) The magnetization curve of a finite-sized spin chain 
for the $S=1/2$ $DH$ model with $L=16$. In addition to the crystal momentum $k$, the eigenvalue $N$ of the ground state is shown. The dotted line shows the position on the magnetic field axis at which the $N=1$ state 
(indicated by the red arrow) is the ground state.  (Right) 
The energy spectrum, i.e., the set of eigenenergy and the crystal momentum of the system with $L=40$. 
The magnetic field is fixed at the value $H=0.831|D|$, where the $N=1$ state is the ground state. 
The ground state is indicated by the red arrow. 
The red and blue points are the eigenstates
for which $N=1$ and $N=2$, respectively. The open green circle is an $N=0$ eigenstate. }
\label{fig: SpectrumS0.5DH}
\end{figure}

The left panel of Fig.~\ref{fig: SpectrumS0.5DH} depicts 
the magnetization curve of finite-sized systems of 
the spin $S=1/2$ $DH$ model. 
Notice that for this model, the ground state is characterized by both the soliton number $N$ and the 
crystal momentum, as is made explicit in this figure.  
It is seen that 
the soliton number $N$ decreases (increases) 
in a stepwise manner with increasing (decreasing) magnetic field. The dotted line indicates the magnetic field
at which 
the single soliton state $N=1$ is the ground state. The low energy sector of the energy spectrum at this value of the magnetic field is shown in the right panel of Fig.~\ref{fig: SpectrumS0.5DH}. The red dots represent the energy eigenstates with $N=1$, where we see two bands of single soliton states. Both bands have their minimum energy at the crystal momentum $k=\pi$. The blue dots form two continua and one isolated branch. The continua correspond to the scattering states of two soliton excitations. We attribute the isolated branch 
to the two soliton bound states formed by the repulsive interaction between the solitons. These figures 
clearly show that the crystal momentum $k$ of the lowest energy state for the sector with the soliton number $N$ is given by $k=\pi N$. We will provide a proof of this generic property in the next subsection.
\ \\
\subsubsection{Exact results}
\begin{figure}
\includegraphics[width=0.7\columnwidth,pagebox=artbox]{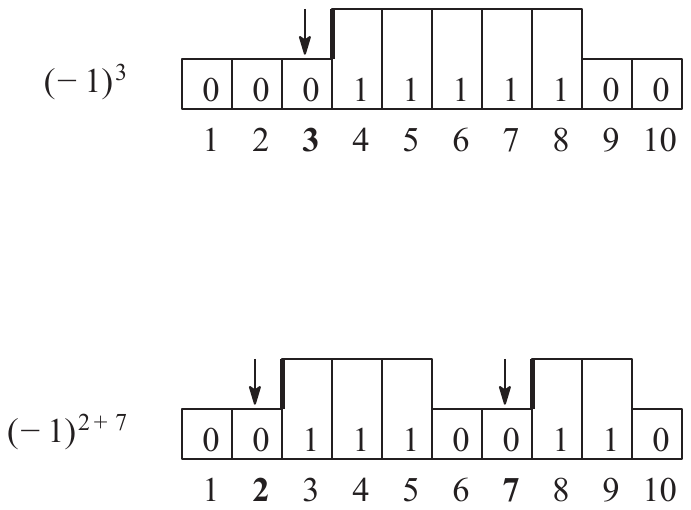}
    \caption{Schematic illustration of the sign $(-1)^{\delta(\bm{n})}$ for $S=1/2$. The symbol $\delta(\bm{n})$ was defined by Eq.~\eqref{eq: sign-def}. }
    \label{fig: fig-sign-example-Sonehalf.pdf}
\end{figure}

\noindent 
{\bf Theorem 1}\\
Consider those eigenstates of the $S=1/2$ $DH$ model Eq.~\eqref{eq: HDH} for which the eigenvalue of $\hat{N}$ is $N$. The crystal momentum of the lowest energy eigenstate is then $k=\pi N$.
\\
\ \\
{\bf Definition: Signed basis}\\
We generate a new set of basis states by 
multiplying each element $\vert {\bm n}\rangle$ of the original basis 
by the factor $(-1)^{\delta(\bm{n})}$, where 
\begin{equation}
\delta(\bm{n})=\sum_{j=1}^L j\left(n_{j+1}-n_{j}+|n_{j+1}-n_{j}|\right)/2.   \label{eq: sign-def}
\end{equation}
This defines our {\it signed basis}. 
We also define the sign of $\bm{n}$ by  $(-1)^{\delta(\bm{n})}$. 
For instance the signed basis states
corresponding to Eqs.~\eqref{eq: N=1example} and \eqref{eq: N=2example} are
\begin{align}
(-1)^3&|00\overset{\downarrow}{0}1111100\rangle,\label{eq: N=1example-signed}
\\
(-1)^{2+7}&|0\overset{\downarrow}{0}1110\overset{\downarrow}{0}110\rangle\label{eq: N=2example-signed},
\end{align}
as depicted in Fig.~\ref{fig: fig-sign-example-Sonehalf.pdf}.
We see that $\delta(\bm{n})$ coincides with the 
sum of the site indices of the right-most 
entry of $^^ ^^ 0"$ (indicated by arrows)
within each segment of the background i.e. consecutive appearances of 0.
In the following we will be using the terminology {\it the site index of a $^^ ^^ 01"$ boundary}. By this we refer to the site index of the $^^ ^^ 0"$ immediately to the left of a 
soliton. Examples of such sites are indicated by arrows in 
Eqs.~\eqref{eq: N=1example-signed}, \eqref{eq: N=2example-signed}, and 
Fig.~\ref{fig: fig-sign-example-Sonehalf.pdf}. 
As a direct consequence of 
the definition of $\delta(\bm{n})$ given in 
Eq.~\eqref{eq: sign-def}, 
\begin{equation}
(-1)^{\delta(\bm{n})}=(-1)^{\delta(T(\bm{n}))}(-1)^N.
\label{eq: sgn-n-Tn}
\end{equation}
\ \\
{\bf Lemma 1 (off-diagonal matrix element)}\\
The off-diagonal matrix elements
of $\hat{\mathcal{H}}_{DH}$ Eq.~\eqref{eq: HDH} in the signed basis $(-1)^{\delta(\bm{n})}|\bm{n}\rangle$ 
for $\bm{n}\in V_N$ are non-positive. \\
\ \\
{\it Proof of Lemma 1}\\
\begin{proof}
Since the Zeeman interaction is diagonal in the signed basis, 
it suffices to show that in this basis 
all off-diagonal elements of $\hat{\mathcal{H}}_{\rm DM}$, defined by Eq.~\eqref{eq: HDM0-1/2}, are non-positive.
The actions of the local Hamiltonian $\hat{h}_i$ Eqs.~\eqref{eq: action-of-hi-a}-\eqref{eq: action-of-hi-d} 
can be summarized
as
\begin{equation}
2\hat{h}_i|\bm{n}\rangle =(n_{i-1}-n_{i+1})|\bar{\bm{n}}^{(i)}\rangle, \label{eq: hi-action-summary}   
\end{equation}
where $\bar{\bm{n}}^{(i)}=(\bar{n}^{(i)}_1\bar{n}^{(i)}_2\cdots \bar{n}^{(i)}_L)$ 
and
\begin{equation}
\bar{n}^{(i)}_j=\left\{
\begin{array}{cc}
n_j,\quad&j\ne i\\
1-n_i,\quad&j=i\\
\end{array}
\right.    
\end{equation}
In Eqs.~\eqref{eq: action-of-hi-a} and \eqref{eq: action-of-hi-b}, where $n_{i-1}-n_{i+1}=-1$, the site indices of the $^^ ^^ 01"$ boundary 
differ by one between 
$\bm{n}$ and $\bar{\bm{n}}^{(i)}$ and 
hence 
$\delta(\bm{n})= \delta(\bar{\bm{n}}^{(i)})\pm 1$. 
Acting on a signed basis then results in 
\begin{equation}
2\hat{h}_i(-1)^{\delta(\bm{n})}|\bm{n}\rangle =|n_{i-1}-n_{i+1}| (-1)^{\delta(\bar{\bm{n}}^{(i)})}|\bar{\bm{n}}^{(i)}\rangle.
\label{eq: hi-action-summary-signed basis}   
\end{equation}
Notice that the negative matrix elements of 
Eqs.~\eqref{eq: action-of-hi-a} and \eqref{eq: action-of-hi-b} have 
now acquired a positive sign upon switching to the signed basis. 
This can be understood as having come from the opposite signs 
that $\bm{n}$ and $\bar{\bm{n}}$ possess. 
The relation Eq.~\eqref{eq: hi-action-summary-signed basis} 
applies as well to 
Eqs.~\eqref{eq: action-of-hi-c} and \eqref{eq: action-of-hi-d}, 
since in this case 
$|n_{i-1}-n_{i+1}|=n_{i-1}-n_{i+1}=1$ and the action of $\hat{h}_i$ does not move the site indices of the $^^ ^^ 01"$ boundary, resulting in $\delta(\bm{n})=\delta(\bar{\bm{n}})$.

It then follows that the off-diagonal matrix elements of $\hat{\mathcal{H}}_{\rm DM}$ Eq.~\eqref{eq: HDM0-1/2} and thus those of $\hat{\mathcal{H}}_{DH}$ are non-positive in the signed basis. 
\end{proof}

\begin{figure}
    \centering
\includegraphics[width=0.2\columnwidth,pagebox=artbox]{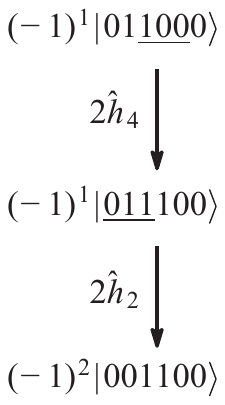}
    \caption{
    Schematic illustration of 
    a 
    one-site translation of a single soliton. The underlined three sites indicate those allowing the 
    outcome of the action of 
    the local Hamiltonian $\hat{h}_i$ 
    on the state to be nonzero.
    }
    \label{fig: fig-hi}
\end{figure}

We consider the states that belong to
Ker($\hat{\mathcal{H}}_{\rm DM}$), i.e., the eigenspace of $\hat{\mathcal{H}}_{\rm DM}$ for zero energy, and other states separately. The basis states for which $|\bm{n}\rangle\in {\rm Ker}(\hat{\mathcal{H}}_{\rm DM})$ satisfy $n_{i}=n_{i+2}$ for all $i\in [1,L]$, which we can confirm by inspection of the action of the DMI on the basis states as discussed in 
Lemma 1. Those $|\bm{n}\rangle$ satisfying this condition belong to one of either spaces:  
$V_0=\{000\cdots 000,111\cdots 111\}$ (the states with $N=0$) or $V_{L/2}=\{101\cdots 010,010\cdots 101\}$ (the states with $N=L/2$). 
Note that these are eigenstates of $\hat{\mathcal{H}}_{DH}$ with eigenenergy $-H\sum_{j=1}^L (S-n_j)$. Among these, the state $|00\cdots00\rangle$ has the lowest energy and 
is a simultaneous eigenstate of $\hat{T}$ and $\hat{N}$ with the eigenvalues $k=0$ and $N=0$. 
Meanwhile the states $|\bm{n}\rangle\notin $ Ker($\hat{\mathcal{H}}_{\rm DM}$) are the eigenstates of $\hat{N}$ with the eigenvalue $N\in [1,L/2-1]$. 
For those states, the following Lemma holds.
\ \\
{\bf Lemma 2 (irreducibility)}\\
For arbitrary pairs of 
$\bm{n}$ and $\bm{n}'$ belonging to $V_N$ with $N\in [1,L/2-1]$, 
there exists a positive integer $l$ such that 
\begin{equation}
(-1)^{\delta(\bm{n})+\delta(\bm{n}')}\langle\bm{n}|(-\hat{\mathcal{H}}_{DH})^l|\bm{n}'\rangle>0.\label{eq: irreducibility-S-1/2}
\end{equation}
\ \\
Though we defer the proof of Lemma 2 to the Supplemental Material\cite{sm}, 
this statement should be intuitively acceptable when one 
realizes
that Eq.~\eqref{eq: irreducibility-S-1/2} holds if 
multiple actions $\hat{h}_{i'}\hat{h}_{i''}\hat{h}_{i'''} \cdots $ of the local Hamiltonians on  state $(-1)^{\delta(\bm{n})}|\bm{n}\rangle$ can translate, shorten, and stretch the solitons in the state. Those operations consist of the one-site shift of the $^^ ^^ 01"$ boundary and $^^ ^^ 10"$ boundary, which can be performed by the local Hamiltonian $\hat{h}_i$ as shown  in Eqs.~\eqref{eq: action-of-hi-a}-\eqref{eq: action-of-hi-d}.
Figure \ref{fig: fig-hi} shows an example of translation of a single soliton by the multiple action of the local Hamiltonians
\begin{equation}
4\hat{h}_{2}\hat{h}_{4}(-1)^1 |011000\rangle  =(-1)^2 |001100\rangle+\mbox{other terms}.  
\end{equation}
The first (the second) step in Fig.~\ref{fig: fig-hi} demonstrates a process where the soliton is stretched (shortened). See the Supplemental Material\cite{sm} for a proof. 
\ \\
\ \\
{\it Proof of Theorem 1}
\begin{proof}
When $N\in [1,L/2-1]$, 
it follows from Lemmas 1 and 2 and the Perron-Frobenius theorem\cite{Tasaki2020} 
that 
(1) the lowest energy eigenstate of $\hat{\mathcal{H}}_{DH}$ 
within 
each eigenspace of $\hat{N}$ is non-degenerate, and (2) the state vector $|E_{{\rm min},N}\rangle$ of the lowest energy state is spanned by the signed basis, where all coefficients are positive, i.e., 
\begin{align}
|E_{{\rm min},N}\rangle=&\sum_{\bm{n}\in V_N}
a(\bm{n})(-1)^{\delta(\bm{n})}|\bm{n}\rangle, 
\label{eq: EminN}
\end{align}
with $a(\bm{n})>0.$
Note that the 
sum over $\bm{n}$ encompasses all $N$ soliton basis states. 
For $N=1$, e.g. we can write $|E_{{\rm min},1}\rangle$ in the form 
\begin{align}
|E_{{\rm min},1}\rangle=&
a_1(-1)^1|010000\rangle+a_2(-1)^2|001000\rangle+\cdots\nonumber\\
+& b_1(-1)^1|011000\rangle+b_2(-1)^2|001100\rangle+\cdots\nonumber\\
+& \cdots
\end{align}
with $a_i,b_i\cdots>0$. Owing to the nondegeneracy of $|E_{{\rm min},N}\rangle$, it is an eigenstate of the site-translation operator $\hat{T}$ and thus 
\begin{equation}
|a(\bm{n})|=|a(T(\bm{n}))|,
\end{equation}
which leads to $a(\bm{n})=a(T(\bm{n}))$ because {\color{black}of the condition} $a(^\forall\bm{n}\in V_N)>0$. We can thus rewrite Eq.~\eqref{eq: EminN} as
\begin{align}
&|E_{{\rm min},N}\rangle\nonumber\\
=&
a(\bm{n})\left((-1)^{\delta(\bm{n})}|\bm{n}\rangle+
(-1)^{\delta(T(\bm{n}))}|T(\bm{n})\rangle+
\cdots\right)\nonumber\\
&+\cdots\nonumber\\
=&
a(\bm{n})(-1)^{\delta(\bm{n})}\left(|\bm{n}\rangle+
(-1)^N|T(\bm{n})\rangle+
\cdots\right)\nonumber\\
&+\cdots\label{eq:  expand}
\end{align}
From the the final expression of the above equation, we see that 
\begin{equation}
\hat{T}|E_{{\rm min},N}\rangle=(-1)^N |E_{{\rm min},N}\rangle
\label{eq: T-eigenvalue}
\end{equation}
for $N\in[1,L/2-1]$. We have already 
shown
that this relation holds for $N=0$ and the states with $N=L/2$ cannot be the ground state. 
This concludes our proof. 
\end{proof}
\section{Higher $S$ model in the limit $J\rightarrow 0$ with finite $D$  }\label{sec: higher s}
\subsubsection{Soliton numbers with various heights and the projected $DH$ model}
For 
$S>1/2$, 
i.e. when $S=1,3/2,2\cdots$, the $DH$ model 
has no conserved quantity. 
However, we have found through finite-sized diagonalization studies that slightly below the critical field, 
a large weight within the ground state wavefunction is dominated by those basis states which can be interpreted as higher spin versions of single solitons.
As an example we display in Eq.~\eqref{eq: dominant-states-S=1} a partial list of such dominant states for the case of $S=1$:
\begin{align}
&|00200000\rangle\nonumber\\
&|00120000\rangle\nonumber\\
&|00210000\rangle\nonumber\\
&|00122000\rangle\nonumber\\
&|02210000\rangle\nonumber\\
&|00211000\rangle\nonumber\\
&\quad\quad\vdots 
\label{eq: dominant-states-S=1}
\end{align}
To fully characterize the wider variety of spatial structures that a soliton can exhibit in the higher $S$ cases, 
we incorporate a set of $2S$ operators $\hat{N}_1,\cdots,\hat{N}_{2S}$ which count the number of solitons of 
various heights,
\begin{align}
\hat{N}_{2S}=
\sum_{i=1}^L \sum_{a<f}\hat{P}_{i-1}^{(S-a)}\hat{P}_{i}^{(S-f)}     ,&\quad f=2S 
\label{eq: Nf-def-f=2S}
\end{align}
and 
\begin{align}
\hat{N}_f=
\sum_{i=1}^L \left( \sum_{a<f}\hat{P}_{i-1}^{(S-a)}-\sum_{b>f}\hat{P}_{i+1}^{(S-b)}\right)\hat{P}_{i}^{(S-f)}     ,&\quad 1\le f<2S. 
\label{eq: Nf-def-f<2S}
\end{align}
In the above we made use of the projection operator 
\begin{equation}
\hat{P}_{i}^{(S-m)}=\prod_{m'\in [0,1,\cdots 2S]/\{m\}}(\hat{S}_i^{z}-S+m')/(m'-m),
\label{eq: P_i-S-f}
\end{equation}
where it is understood that the product over $m'$-values excludes the case $m'=m$.  
When acted on 
a basis vector, this operator yields
\begin{align}
\hat{P}_{i}^{(S-m)}|n_1 n_2 \cdots n_L\rangle
=\delta_{n_i,m}|n_1 n_2 \cdots n_L\rangle.
\end{align}
To familiarize ourselves with how the operators appearing in Eqs.~\eqref{eq: Nf-def-f=2S} and \eqref{eq: Nf-def-f<2S} work, 
it is useful to look into examples of simultaneous eigenstates of multiple $\hat{N}_f$'s.  
We denote the eigenvalue of $\hat{N}_f$ by $N_f$, and begin with the case $S=1$. The states shown in Eq.~\eqref{eq: dominant-states-S=1} are those for which  $(N_1,N_2)=(0,1)$. Below we provide examples of states with a different set of $(N_1 , N_2 )$ values   
\begin{subequations}
\begin{align}
&|00000000\rangle,\quad (N_1,N_2)=(0,0)\label{eq: examples-states-S=1-a}\\
&|11111111\rangle,\quad (N_1,N_2)=(0,0)\label{eq: examples-states-S=1-b}\\
&|22222222\rangle,\quad (N_1,N_2)=(0,0)\label{eq: examples-states-S=1-c}\\
&|00100000\rangle,\quad (N_1,N_2)=(1,0)\label{eq: examples-states-S=1-d}\\
&|00111111\rangle,\quad (N_1,N_2)=(1,0)\label{eq: examples-states-S=1-e}\\
&|00100211\rangle,\quad (N_1,N_2)=(1,1)\label{eq: examples-states-S=1-f}\\
&|00100110\rangle,\quad (N_1,N_2)=(2,0)\label{eq: examples-states-S=1-g}\\
&|01200210\rangle,\quad (N_1,N_2)=(0,2)\label{eq: examples-states-S=1-h}\\
&|11121111\rangle,\quad (N_1,N_2)=(-1,1)\label{eq: examples-states-S=1-i}\\
&|11212211\rangle,\quad (N_1,N_2)=(-2,2)\label{eq: examples-states-S=1-j}.
\end{align}
\end{subequations}
The states taken up in Eqs.~\eqref{eq: examples-states-S=1-a}-\eqref{eq: examples-states-S=1-c} contain no solitons. 
Inspection of Eqs.~\eqref{eq: dominant-states-S=1} and Eqs.~\eqref{eq: examples-states-S=1-d}-\eqref{eq: examples-states-S=1-h} 
reveals that for these cases, $N_1$ ($N_2$) represents the number of segments 
consisting of consecutive ^^ ^^ 1"s (^^ ^^ 2"s) and forming a local maximum. Finally equations Eqs.~\eqref{eq: examples-states-S=1-i} and \eqref{eq: examples-states-S=1-j} show that negative $N_1$ represents the number of segments made up of consecutive ^^ ^^ 1"s which form a local minimum.   

We next turn to examples of states for $S=3/2$ 
\begin{subequations}
\begin{align}
&|00100000\rangle,\quad (N_1,N_2,N_3)=(1,0,0)\label{eq: examples-states-S=3/2-d}\\
&|01211000\rangle,\quad (N_1,N_2,N_3)=(0,1,0)\label{eq: examples-states-S=3/2-f}\\
&|01123100\rangle,\quad (N_1,N_2,N_3)=(0,0,1)\label{eq: examples-states-S=3/2-g}\\
&|01310230\rangle,\quad (N_1,N_2,N_3)=(0,0,2)\label{eq: examples-states-S=3/2-h}\\
&|11131111\rangle,\quad (N_1,N_2,N_3)=(-1,0,1)\label{eq: examples-states-S=3/2-i}\\
&|22232222\rangle,\quad (N_1,N_2,N_3)=(0,-1,1).\label{eq: examples-states-S=3/2-j}
\end{align}
\end{subequations}
Equations \eqref{eq: examples-states-S=3/2-d} 
and \eqref{eq: examples-states-S=3/2-f} serve to illustrate that $(N_1,N_2,0)$ for $S=3/2$ coincides with the state with $(N_1,N_2)$ for $S=1$.
Meanwhile from Eqs.~\eqref{eq: examples-states-S=3/2-g} and ~\eqref{eq: examples-states-S=3/2-h} we 
see that for examples like these, $N_3$ gives the number of segments made up of consecutive ^^ ^^ 3"s,  forming a local maximum.
Equations \eqref{eq: examples-states-S=3/2-i} and \eqref{eq: examples-states-S=3/2-j} tell us that a negative $N_1$ $(N_2)$ counts the number of segments consisting of consecutive ^^ ^^ 1"s (^^ ^^ 2"s),  each forming a local minimum. 
Based on these observations, we regard a positive $N_f$ as indicating the number of {\it solitons} of height (amplitude) $f$, and negative $N_f$ the number of {\it valleys} of depth $f$. 

To make further progress, we shall assume that the essential properties of the ground state, such as its crystal momentum, remains intact  even if we truncate the matrix elements in the Hamiltonian $\hat{\mathcal{H}}_{DH}$ connecting sectors 
belonging to different eigenvalue sets $N_1,\cdots, N_{2S}$. Before going further we will first provide a numerical check on the plausibility 
of this working assumption. 

\begin{figure}
\begin{center}
\includegraphics[width=\columnwidth,pagebox=artbox]{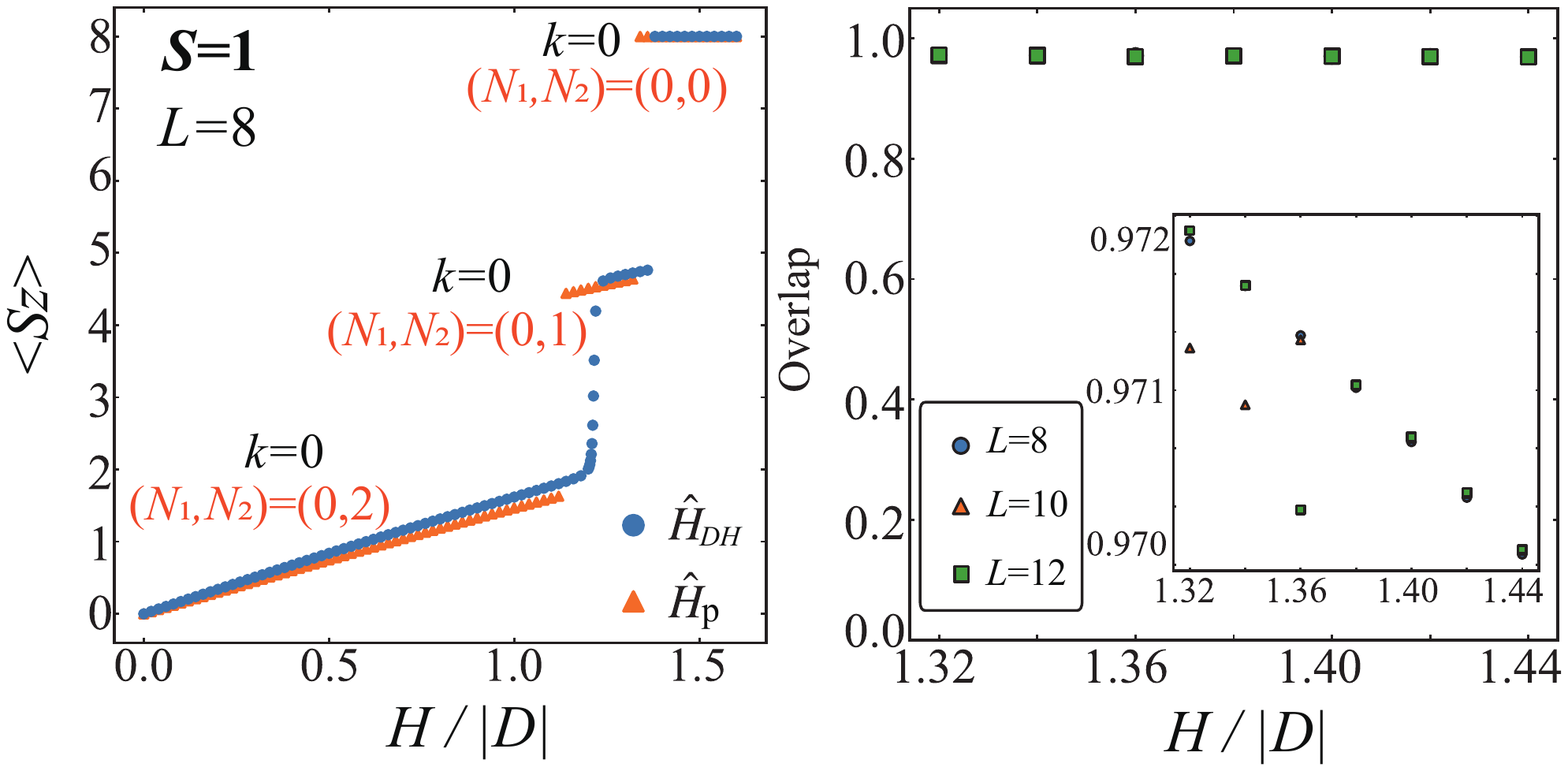}
\end{center}
\caption{Left panel: 
Magnetization curves for the two $S=1$ models 
$\hat{\mathcal{H}}_{DH}$ and $\hat{\mathcal{H}}_{{\rm p}}$. The system size is set at $L=8$. 
The crystal momentum in the ground state is always zero for 
both models. The number of solitons $N_f$ of heights $f=1,2$ present in the ground state $|{\rm g}_{\rm p}\rangle$ for $\hat{\mathcal{H}}_{{\rm p}}$ is also shown. 
Right panel: 
Overlap $|\langle {\rm g}_{\rm p}|{\rm g}_{DH}\rangle|$ between the ground states $|{\rm g}_{\rm p}\rangle,|{\rm g}_{DH}\rangle$ for the two models for the system sizes $L=8$, 10, 12. 
The magnetic field lies in the range $H/D=1.32-1.44$, where single soliton states of maximal height (=2) dominate the ground state of $\hat{\mathcal{H}}_{DH}$.  
The inset of the right panel is a blowup of the vertical axis.}
\label{fig: OverlapS1_HDHvsHM}
\end{figure}

\begin{figure}
\begin{center}
\includegraphics[width=\columnwidth,pagebox=artbox]{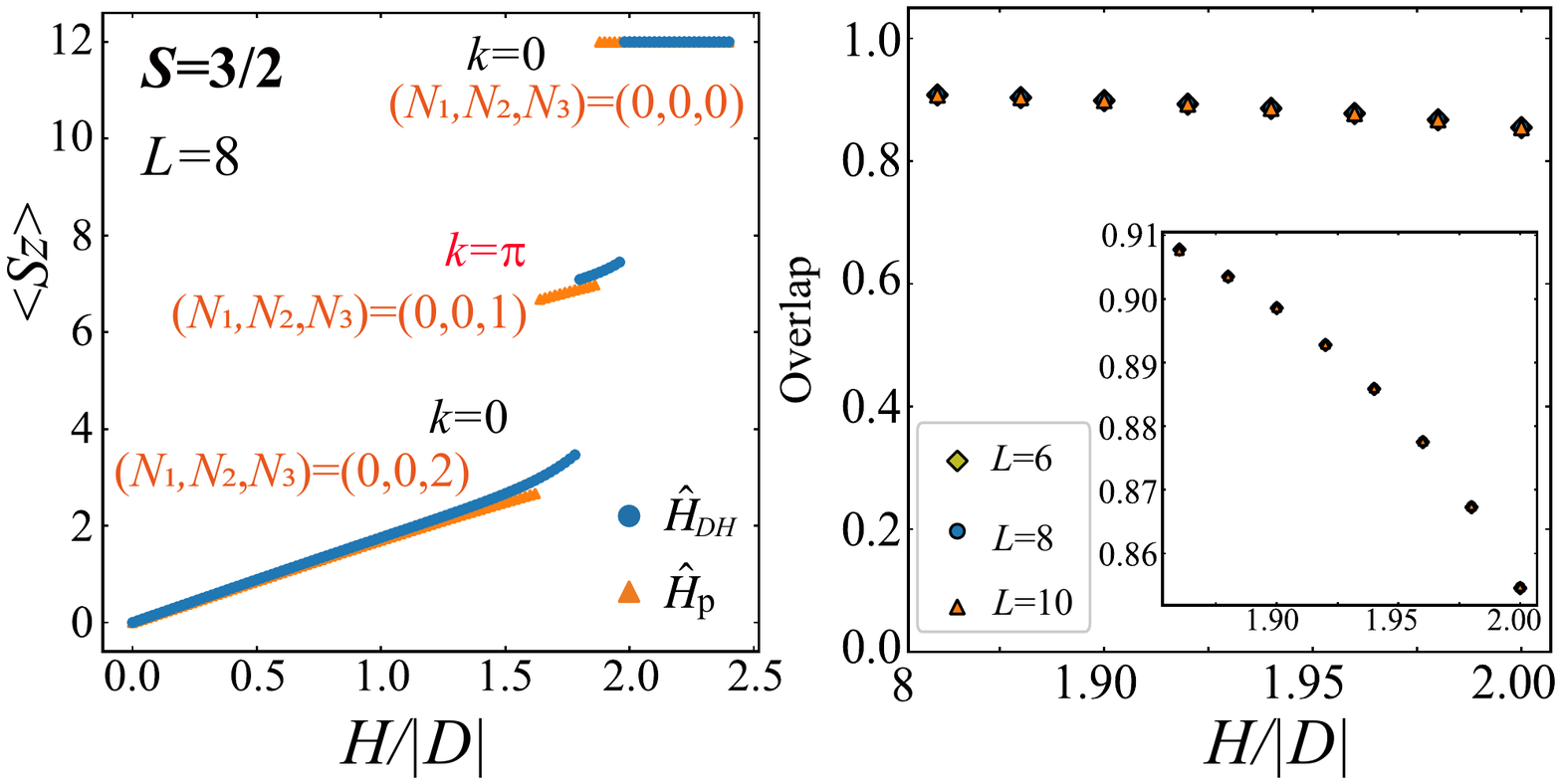}
\end{center}
\caption{Left panel: Magnetization curves for $S=3/2$ $\hat{\mathcal{H}}_{DH}$ and $\hat{\mathcal{H}}_{{\rm p}}$ for 
finite-sized systems with $L=8$. The crystal momentum in the ground state changes by $\pm \pi$ 
accompanying the discontinuous changes in the magnetization. 
The number of solitons $N_f$ of heights $f=1,2$ present in the ground state $|{\rm g}_{\rm p}\rangle$ for $\hat{\mathcal{H}}_{{\rm p}}$ is also shown. 
Right panel: Overlap $|\langle {\rm g}_{\rm p}|{\rm g}_{DH}\rangle|$ between the ground states $|{\rm g}_{\rm p}\rangle,|{\rm g}_{DH}\rangle$ for the two models for $S=1$ with $L$=8, 10, 12 for the magnetic field in the range of $H/D=1.86-2.00$, where the single soliton states with maximum height (=3 for $S=3/2$) dominate the ground state of $\hat{\mathcal{H}}_{DH}$. The inset of the right panel is a blowup of the vertical axis.}
\label{fig: OverlapS1.5_HDHvsHM}
\end{figure}

Let $\hat{P}(\{N_f\})$ 
then 
be the projection operator into the eigenspace with $N_1,\cdots,N_{2S}$. We introduce the truncated Hamiltonian 
\begin{equation}
\hat{\mathcal{H}}_{{\rm p}}=\sum_{N_1,\cdots, N_{2S}} \hat{P}(\{N_f\})\hat{\mathcal{H}}_{DH}\hat{P}(\{N_f\}),   \label{eq: H_M}
\end{equation}
which we shall call the projected $DH$ (p$DH$) model.   
Figures~\ref{fig: OverlapS1_HDHvsHM} and \ref{fig: OverlapS1.5_HDHvsHM} are numerical results for finite-sized systems for the cases $S=1,3/2$, and  
demonstrate that the series of states containing solitons of maximal height, 
which can be categorized as 
$(N_1,N_2)=(0,N_s)$ and $(N_1,N_2,N_3)=(0,0,N_s)$ with $N_s=0, 
1, 2, \cdots$ dominate over other states in the magnetization process in the DH model. 
This explains why the basic features of the magnetization curves for $\hat{\mathcal{H}}_{DH}$ 
are similar to those for $\hat{\mathcal{H}}_{{\rm p}}$.
We 
have confirmed that the overlaps between the ground states for $\hat{\mathcal{H}}_{DH}$ and $\hat{\mathcal{H}}_{{\rm p}}$ slightly below 
their respective critical fields are larger than 97$\%$ for $S=1$, and larger than 91$\%$ for $S=3/2$. 

Having thus seen that it is reasonable for our purpose to work with $\hat{\mathcal{H}}_{{\rm p}}$, we turn to its spectral properties that can be established in a rigorous manner. 
\subsubsection{Exact results}
\begin{figure}
\includegraphics[width=0.8\columnwidth,pagebox=artbox]{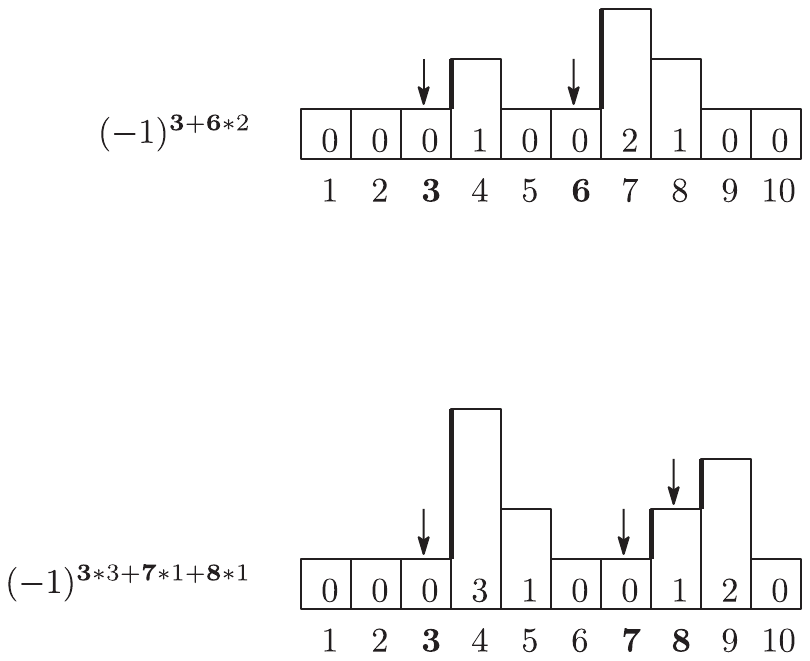}
    \caption{
    Schematic illustration of the sign $(-1)^{\delta(\bm{n})}$ for $S=1$ (upper panel) and $S=3/2$ (lower panel). The symbol $\delta(\bm{n})$ was defined by Eq.~\eqref{eq: sign-def}.
    }
    \label{fig: fig-higherS}
\end{figure}

\ \\
{\bf Theorem 2 (height parity effect)}\\
The lowest energy eigenstate of the p$DH$ Hamiltonian $\hat{\mathcal{H}}_{{\rm p}}$
(Eq.~\eqref{eq: H_M}),  
within the sector where the eigenvalues of the operators $\hat{N}_1,\hat{N}_2,\cdots ,\hat{N}_{2S}$ are 
$N_1,N_2,\cdots,N_{2S}$, respectively, has the crystal momentum $k=\pi \sum_{f=1}^{2S} f N_f$. \\
As a consequence of this theorem, we find that\\ 
{\bf Corollary 1 (spin parity effect)}\\
The lowest energy eigenstate of the p$DH$ Hamiltonian $\hat{\mathcal{H}}_{{\rm p}}$,  
within the sector where the eigenvalues of the operators $\hat{N}_1,\hat{N}_2,\cdots ,\hat{N}_{2S}$ are 
$0,0,\cdots,N_{2S}$, respectively, has the crystal momentum $k=2\pi S N_{2S}$. \\
\ \\
This corollary, encompassing Theorem 1 which is specific to $S=1/2$, and its higher $S$ generalizations, will later be seen to be of direct relevance in understanding the magnetization behavior of the models 
discussed in this paper.
As with   
Theorem 1, 
a crucial part of proving Theorem 2 consists in finding 
a signed basis such that the off-diagonal matrix element of 
$\hat{\mathcal{H}}_{{\rm p}}$ is non-positive. 
For that purpose 
we introduce a unitary operator
\begin{equation}
    \hat{U}:=\exp\left[i\pi \sum_{j=1}^L\sum_{f=1}^{2S}\sum_{a=0}^{2S-f}j f \hat{P}_{j}^{(S-a)}\hat{P}_{j+1}^{(S-a-f)}\right].
\label{def of U_M}
\end{equation}
We observe that the projection operator $\hat{P}_{j}^{(S-m)}$ as defined in Eq.~\eqref{eq: P_i-S-f}, and therefore the unitary operator $\hat{U}$
can be expressed solely in terms of  \{$\hat{S}_i^z$\}.
\noindent
 Thus the basis vectors $|\bm{n}\rangle$ are 
eigenvectors of $\hat{U}$, where it is clear from 
the definition Eq.~\eqref{def of U_M} that the corresponding 
eigenvalues are sign factors dependent on $\bm{n}$. One can 
verify that this dependence can be written explicitly 
in the following form: 
\begin{equation}
\hat{U}|\bm{n}\rangle
=(-1)^{\delta(\bm{n})}|\bm{n}\rangle,
\end{equation}
where $\delta(\bm{n})$ was defined back in Eq.~\eqref{eq: sign-def} when 
dealing with the $S=1/2$ case. We note that the integer 
$n_i$ which appears in Eq.~\eqref{eq: sign-def} now take the 
values $\{0,1,2,\cdots,2S\}$.
Figure \ref{fig: fig-higherS} shows examples of $\delta(\bm{n})$ for higher $S$. 
\ 

We also introduce the notation $V(\{N_f\})$ as the set of $\bm{n}$ such that 
$|\bm{n}\rangle$ is a simultaneous eigenvector of $\hat{N}_1,\cdots,\hat{N}_{2S}$ with eigenvalue $N_1,\cdots,N_{2S}$. We also denote by $V_\mu(\{N_f\})$ the subset of $V(\{N_f\})$ in which the Hamiltonian $\hat{\mathcal{H}}_{{\rm p}}$ is irreducible. The index $\mu$ runs %
from 1 to $\sharp(\{V_\mu\{N_f\}\}_\mu)$, which is the number of irreducible subspaces in $V(\{N_f\})$. We denote the dimension of $V_\mu(\{N_f\})$ by $d( V_\mu(\{N_f\})$.

The proof of %
Theorem 2 relies on the following four Lemmas.
\ \\
{\bf Lemma 3 (sign of translated basis states)}\\
\begin{equation}
(-1)^{\delta(\bm{n})}=(-1)^{\delta(T(\bm{n}))}(-1)^{\sum_{f=1}^{2S}f N_f},
\label{eq: sgn-n-Tn-higherS}
\end{equation}
which is a generalization of Eq.~\eqref{eq: sgn-n-Tn} to higher $S$. 

The proof of Lemma 3 will be given after Lemma 6 is stated. 
\ \\
{\bf Lemma 4 (off-diagonal matrix element)}\\
In the signed basis, the off-diagonal matrix elements are non-positive, i.e.
\begin{equation}
(-1)^{\delta(\bm{n}')+\delta(\bm{n})}\langle \bm{n}'|
\hat{\mathcal{H}}_{{\rm p}}|\bm{n}\rangle\le 0, \end{equation}
which is a generalization of Lemma 1 to higher $S$. 

The proof of Lemma 4 will be given after Lemma 6 is stated. 
\ \\
{\bf Lemma 5 (\bf kernel of $\hat{\mathcal{H}}_{{\rm p}}(H=0)$)}\\
Let us denote by $\hat{\mathcal{H}}_{{\rm p}0}$ the Hamiltonian $\hat{\mathcal{H}}_{{\rm p}}(H=0)$.
Among the states in Ker($\hat{\mathcal{H}}_{\rm p0}$), the state $|00\cdots00\rangle$ has the lowest energy and is a simultaneous eigenstate of $\hat{T}$ and $\hat{N}$ with $k=0$ and $N_1=N_2=\cdots=N_{2S}=0$.

The proof of Lemma 5 will be given after Lemma 6 is stated. 
\ \\
{\bf Lemma 6 (\bf irreducibility)}\\
When $|\bm{n}\rangle\in V_\mu(\{N_f\})$ with $d(V_\mu(\{N_f\}))>1$, $|T(\bm{n})\rangle\in V_\mu(\{N_f\})$, i.e. there is a positive integer $l$ such that
\begin{equation}
\langle \bm{n}|(\hat{\mathcal{H}}_{{\rm p}})^l|T(\bm{n})\rangle\ne 0\label{eq: statement-of-Lemma6}
\end{equation}
We defer the proof of Lemma 6 to the Supplemental Material\cite{sm}.\\
\ \\
{\it Proof of Lemma 3}
\begin{proof}
We begin by observing that the site-translation operator $\hat{T}$ commutes with $\hat{N}_f$ for $f=1,\cdots ,2S$ under the periodic boundary condition $\hat{\bm{S}}_{L+1}=\hat{\bm{S}}_{1}$, because the site-translation does not change the numbers and heights (depths) of solitons (valleys). A straightforward calculation shows that 
$\hat{T}$ is transformed via a unitary operator
as 
\begin{equation}
\hat{U}\hat{T}=\exp\left(i\pi \sum_{f=1}^{2S}f \hat{N}_f\right)\hat{T}\hat{U}.
\label{eq: UTU}
\end{equation}
The derivation of Eq.~\eqref{eq: UTU} is deferred to the Supplemental Material\cite{sm}. 
Using the definition of $|T(\bm{n})\rangle$ and taking into account the sign in Eq.~\eqref{eq: sgn-n-Tn-higherS}, we see that 
\begin{equation}
\hat{U}\hat{T}|\bm{n}\rangle=(-1)^{\delta(T(\bm{n}))}|T(\bm{n})\rangle   
\label{eq: signTn}
\end{equation}
can be rewritten as
\begin{align}
\hat{U}\hat{T}|\bm{n}\rangle&=\exp\left(i\pi \sum_{f=1}^{2S}f \hat{N}_f\right)\hat{T}\underbrace{\hat{U}|\bm{n}\rangle}_{(-1)^{\delta(\bm{n})}|\bm{n}\rangle}\nonumber\\
&=(-1)^{\delta(\bm{n})}\exp(i\pi \sum_{f=1}^{2S}f \hat{N}_f)\hat{T}|\bm{n}\rangle\nonumber\\
&=(-1)^{\delta(\bm{n})}\hat{T}\exp(i\pi \sum_{f=1}^{2S}f \hat{N}_f)|\bm{n}\rangle\nonumber\\
&=(-1)^{\delta(\bm{n})+\sum_{f=1}^{2S}f N_f}\hat{T}|\bm{n}\rangle\nonumber\\
&=(-1)^{\delta(\bm{n})+\sum_{f=1}^{2S}f N_f}|T(\bm{n})\rangle.\label{eq: UTn-ket}
\end{align}
Equating the right-hand side of Eq.~\eqref{eq: signTn}
with the expression given in the last line of Eq.~\eqref{eq: UTn-ket}, 
we arrive at Eq.~\eqref{eq: sgn-n-Tn-higherS}. 
\end{proof}
\ \\
{\it Proof of Lemma 4}
\begin{proof}
Let us start by recasting the Hamiltonian of Eq.~\eqref{eq: H_M} into a form more suitable   
for the present discussion. 
Consider restricting the local Hamiltonian to the Hilbert space spanned by $|\bm{n}\rangle$ satisfying
\begin{equation}
{\rm min}(n_{i-1},n_{i+1})\le n_i\le {\rm max}(n_{i-1},n_{i+1}).\label{eq: restriction-condition}  
\end{equation}
The numbers $\{\hat{N}_f\}_{f=1}^{2S}$ are then conserved. 
In contrast, $\hat{h}_i$ changes the height of a soliton with a peak at $i$ if ${\rm max}(n_{i-1},n_{i+1}) < n_i$. 
Likewise when $n_i<{\rm min}(n_{i-1},n_{i+1})$, $\hat{h}_i$ changes the depth of its valley 
whose lowest point resides at site $i$. 
Introduce a projection operator onto the space satisfying \eqref{eq: restriction-condition} as
\begin{equation}
    \hat{\mathcal{P}}_i =\sum_{a=0}^{2S}\sum_{b=0}^{2S}
    \hat{P}_{i-1}^{(S-a)}\hat{P}_{i+1}^{(S-b)}
\left(\sum_{k={\rm min}(a,b)}^{{\rm max}(a,b)}\hat{P}_i ^{(S-k)}\right). 
\end{equation}
In terms of $\hat{\mathcal{P}}_i$, $\hat{\mathcal{H}}_{{\rm p}}$ can be rewritten as
\begin{equation}
\hat{\mathcal{H}}_{{\rm p}}=-H\sum_{i=1}^L \hat{S}_i^z
-\frac{D}{2}\sum_{i=1}^L \hat{\mathcal{P}}_i\hat{h}_i\hat{\mathcal{P}}_i,  \end{equation}
where $\hat{h}_i$ was defined by Eq.~\eqref{eq: hi}. 
The summand in the second term in the right-hand side is further rewritten as
\begin{equation}
\hat{\mathcal{P}}_i\hat{h}_i\hat{\mathcal{P}}_i
=2
\sum_{a=0}^{2S}\sum_{b=0}^{2S}
(a-b)\hat{\mathcal{S}}_i^x (a,b)
\hat{P}_{i-1}^{(S-a)}\hat{P}_{i+1}^{(S-b)},\label{eq: Pi-hi-Pi}
\end{equation}
with  
\begin{equation}
\hat{\mathcal{S}}_i^x (a,b):=
\left(\sum_{k={\rm min}(a,b)}^{{\rm max}(a,b)}\hat{P}_i ^{(S-k)}\right)
\hat{S}_i^x 
\left(\sum_{k'={\rm min}(a,b)}^{{\rm max}(a,b)}\hat{P}_i ^{(S-k')}\right).\label{eq: mathcalSix}
\end{equation}
Under the unitary tranformation 
\begin{align}
&\hat{\mathcal{H}}_{{\rm p}}':=
\hat{U}\hat{\mathcal{H}}_{{\rm p}}\hat{U}^\dagger,
\end{align}
our Hamiltonian $\hat{\mathcal{H}}_{{\rm p}}$
becomes 
\begin{equation}
\hat{\mathcal{H}}_{{\rm p}}'=
-H\sum_{i=1}^L \hat{S}_i^z
-D\sum_{i=1}^L \hat{h}'_i,
\end{equation}
where
\begin{align}
\hat{h}'_i
&=\sum_{a=0}^{2S}\sum_{b=0}^{2S}
(a-b)\hat{U}\hat{\mathcal{S}}_i^x (a,b)
\hat{P}_{i-1}^{(S-a)}\hat{P}_{i+1}^{(S-b)}\hat{U}^\dagger\label{eq:  hprimei-first-line}\\
&=
\sum_{a=0}^{2S}\sum_{b=0}^{2S}
|a-b|\hat{\mathcal{S}}_i^x (a,b)
\hat{P}_{i-1}^{(S-a)}\hat{P}_{i+1}^{(S-b)}.\label{eq: hprimei}
\end{align}
Details on how the first line of the above equation, Eq.~\eqref{eq:  hprimei-first-line}, leads to the second, Eq.~\eqref{eq: hprimei}, is provided in the Supplemental Material\cite{sm}.
Noting that $\hat{\mathcal{S}}_i^x (a,b)$ is the product of projection operators and $\hat{S}_i^x$  
leads to 
\begin{equation}
\langle \bm{n}'|
\hat{\mathcal{H}}'_{{\rm p}}|\bm{n}\rangle\le 0,\quad \mbox{ for }\bm{n}\ne \bm{n}'.  
\end{equation}
Using this relation, we find that 
\begin{align}
&(-1)^{\delta(\bm{n}')+\delta(\bm{n})}\langle \bm{n}'|
\hat{\mathcal{H}}_{{\rm p}}
|\bm{n}\rangle\nonumber\\
&=\langle \bm{n}'|\hat{U}\hat{\mathcal{H}}_{{\rm p}}\hat{U}^\dagger|\bm{n}\rangle\nonumber\\
&=\langle \bm{n}'|
\hat{\mathcal{H}}'_{{\rm p}}
|\bm{n}\rangle\le 0. 
\end{align}
\end{proof}
\ \\
{\it Proof of Lemma 5}
\begin{proof}
The set of the states $|\bm{n}\rangle\in V_\mu(\{N_f\})$ with $d(V_\mu(\{N_f\}))=1$ forms the basis of Ker($\hat{\mathcal{H}}_{\rm p0}$). Each basis vector $|\bm{n}\rangle$ is an eigenstate of the Zeeman energy. The fully polarized state $|00\cdots00\rangle$ has the lowest Zeeman energy and thus is the lowest energy state in  Ker($\hat{\mathcal{H}}_{\rm p0}$). 
\end{proof}
\ \\
{\it Proof of Theorem 2}\\
This proof is similar to that for 
Theorem 1, and proceeds by replacing $(-1)^N$ by  $(-1)^{\sum_{f=1}^{2S} f N_f}$ in Eqs.~\eqref{eq: sgn-n-Tn}, \eqref{eq: expand}, and \eqref{eq: T-eigenvalue} 
.
\begin{proof}
When the ground state is given by $|00\cdots00\rangle$, 
theorem 2 holds (Lemma 5). According to Lemma 5, other $|{\bm n}\rangle$s belonging to $V_{\mu}(\{
N_f \})$ with $d(V_{\mu}(\{N_f \}))=1$ cannot be the ground state. We thus focus on the case where the ground state is spanned by 
$|{\bm n}\rangle$ with ${\bm n}$ 
belonging to $V_{\mu}(\{N_f \})$ for which $d(V_{\mu}(\{N_f \})) > 1$.

When $N_f\ne 0$ for a certain $f$, 
it follows from Lemmas 4 and 6 and the Perron-Frobenius theorem\cite{Tasaki2020} that
the lowest energy eigenstate of $\hat{\mathcal{H}}_{{\rm p}}$ for each eigenspace 
consisting of $\bm{n}\in V_\mu(\{N_f\})$ is non-degenerate and the state vector $|E_{{\rm min},\{N_f\},\mu}\rangle$ is spanned by the signed basis 
where all coefficients are positive,
i.e.,
\begin{align}
|E_{{\rm min},\{N_f\},\mu}\rangle=&\sum_{\bm{n}\in V_\mu(\{N_f\})}
a(\bm{n})(-1)^{\delta(\bm{n})}|\bm{n}\rangle,
\label{eq: EminNf}
\end{align}
with $a(\bm{n})>0$. 
From the nondegeneracy of $|E_{{\rm min},N}\rangle$, it is an eigenstate of the site-translation operator $\hat{T}$ and thus 
\begin{equation}
|a(\bm{n})|=|a(T(\bm{n}))|,
\end{equation}
which leads to $a(\bm{n})=a(T(\bm{n}))$ because $a(^\forall\bm{n}\in V_N)>0$. We can thus rewrite Eq.~\eqref{eq: EminNf} as
\begin{align}
&|E_{{\rm min},\{N_f\},\mu}\rangle\nonumber\\
=&
a(\bm{n})\left((-1)^{\delta(\bm{n})}|\bm{n}\rangle+
(-1)^{\delta(T(\bm{n}))}|T(\bm{n})\rangle+
\cdots\right)\nonumber\\
&+\cdots\label{eq: Prop3 used here}\\
=&
a(\bm{n})(-1)^{\delta(\bm{n})}\left(|\bm{n}\rangle+
(-1)^{\sum_{f=1}^{2S} f N_f}|T(\bm{n})\rangle+
\cdots\right)\nonumber\\
&+\cdots.\label{eq:  expand-Nf}
\end{align}

This implies that  
\begin{equation}
\hat{T}|E_{{\rm min},\{N_f\},\mu}\rangle=(-1)^{\sum_{f=1}^{2S} f N_f}
|E_{{\rm min},\{N_f\},\mu}\rangle.
\label{eq: T-eigenvalue-Nf}
\end{equation}
\end{proof}
\section{Chirality of Quantum Solitons\label{sec: chirality}}
\begin{center}
\begin{figure}
\includegraphics[width=0.8\columnwidth,pagebox=artbox]{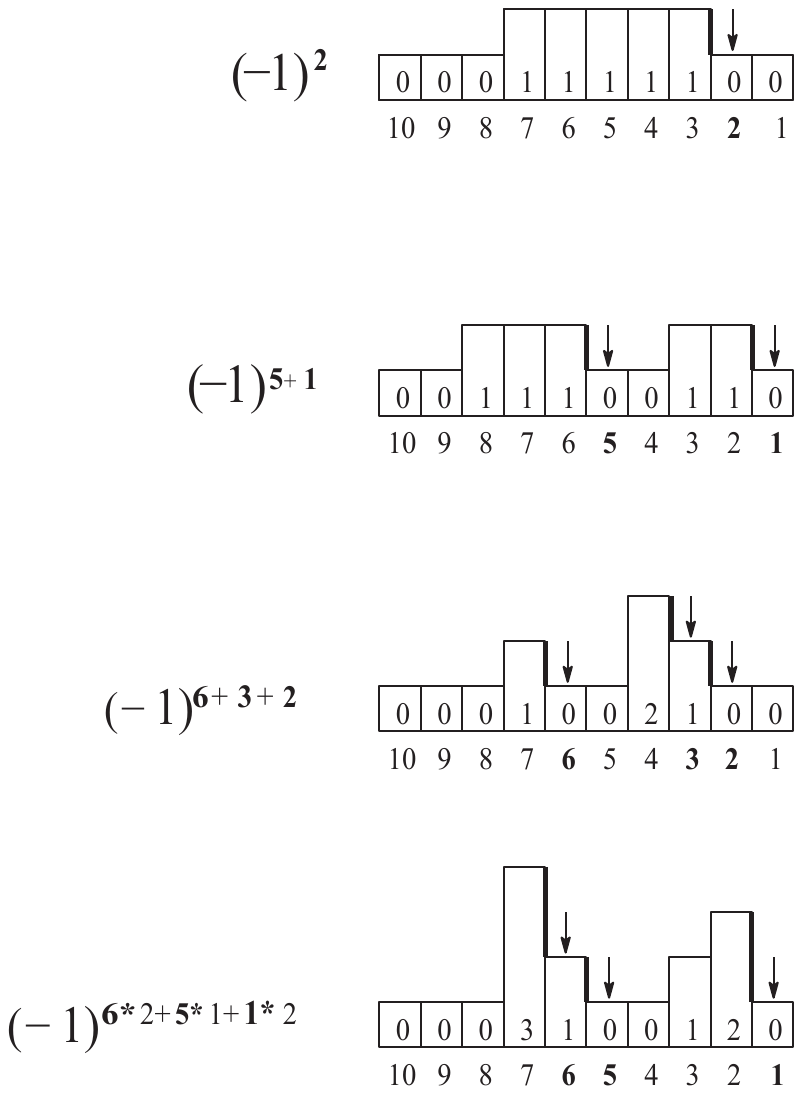}
\caption{Examples of signed basis for negative $D$. The upper two panels represent basis states for $S=1/2$ corresponding to those shown in Fig.~\ref{fig: fig-sign-example-Sonehalf.pdf}. The sign $(-1)^{\delta^{(-)}(\bm{n})}$ uses $\delta^{(-)}(\bm{n})$ defined in Eq.~\eqref{eq: EminN-D<0}. 
The lower two panels represent basis states for $S=1$ and $S = 3/2$, which correspond to the figures shown in Fig.~\ref{fig: fig-higherS}. The numbers underneath the figures represent $L-j$ %
where $j$ is the site index.}
\label{fig: negative-D}
\end{figure}
\end{center}

Up to now we have not explicitly addressed the role that chirality plays in our problem. Let us define the 
chirality of a state by 
\begin{equation}
\mbox{ch}=\langle 
\sum_j \left(\hat{\bm{S}}_j \times\hat{\bm{S}}_{j+1}\right)_y\rangle=\frac{1}{D}\langle\hat{\mathcal{H}}_{\rm DM}\rangle
\label{eq: ch-def}
\end{equation}
in analogy with 
\cite{braunChiralQuantumSpin1996}. 
We can then readily show 
\begin{subequations}
\begin{align}
\mbox{ch}<0,\quad\mbox{when }& D>0\label{eq: ch-D-a}\\    
\mbox{ch}>0,\quad\mbox{when }& D<0\label{eq: ch-D-b}        
\end{align}
\end{subequations}
to be true in the 
lowest energy state in  
any given 
irreducible space $V_\mu(\{N_f\})$ with $d_\mu(\{N_f\})\ne 1$. 
To establish Eq.~\eqref{eq: ch-D-a}, we observe that 
\begin{align}
\mbox{ch}=&\frac{1}{D}\langle E_{{\rm min},\{N_f\},\mu}|\hat{\mathcal{H}}_{\rm DM}|E_{{\rm min},\{N_f\},\mu}\rangle\nonumber\\
=&\frac{1}{D}\sum_{\bm{n},\bm{n}'\in V_\mu(\{N_f\})}
\underbrace{\langle \bm{n}|\hat{\mathcal{H}}_{\rm DM}|\bm{n}'\rangle (-1)^{\delta(\bm{n})+\delta(\bm{n}')}}_{<0}\underbrace{a(\bm{n})a(\bm{n}')}_{>0}\nonumber\\
&<0,
\label{eq: ch-positive-D}    
\end{align}
with the use of Eq.~\eqref{eq: EminN} and theorem 2. 
Equation~\eqref{eq: ch-D-b} can %
likewise be verified. 
In the previous sections, we have always assumed that $D>0$. 
Taking the choice $D<0$ require us to 
replace $\delta(\bm{n})$ with 
\begin{equation}
\delta^{(-)}(\bm{n})=\sum_{j=1}j(n_{L-j}-n_{L+1-j}+|n_{L-j}-n_{L+1-j}|)/2    
\label{eq: sign-D<0-def}
\end{equation}
and Eq.~\eqref{eq: EminN} with 
\begin{align}
|E_{{\rm min},N}\rangle=&\sum_{\bm{n}\in V_N}
a(\bm{n})(-1)^{\delta^{(-)}(\bm{n})}|\bm{n}\rangle.
\label{eq: EminN-D<0}
\end{align}
Equation~\eqref{eq: sign-D<0-def} is obtained by replacing $n_{j}$ with $n_{L+1-j}$ in Eq.~\eqref{eq: sign-def}. Examples of the signed basis for $D<0$ are shown in Fig.~\ref{fig: negative-D}. 
\section{Effect of exchange interactions\label{sec: J}}
Among the exchange interactions in $\hat{\mathcal{H}}_{\rm ch}$, the presence of the Ising term $-J\sum_{i}\hat{S}_i^z\hat{S}_{i+1}^z$ is %
immaterial 
to our argument in the previous section in the sense that it commutes with $\hat{N}$ and it is diagonal in the basis $\{|\bm{n}\rangle\}$. 
The XY term $-J\sum_{i}(\hat{S}_i^x\hat{S}_{i+1}^x +\hat{S}_i^y\hat{S}_{i+1}^y)=\hat{\mathcal{H}}_{\rm XY}$, in contrast, does not conserve $\hat{N}$, e.g.,  
\begin{align}
&\hat{\mathcal{H}}_{\rm XY}|011100\rangle
=-\frac{J}{2}\left(|101100\rangle+|011010\rangle\right),
\label{eq: me_between_N1_N2}
\end{align}
for $S=1/2$. Equation \eqref{eq: me_between_N1_N2} represents the matrix elements between $N=1$ and $N=2$. 
The ground state of $\hat{\mathcal{H}}_{\rm ch}$ is thus given by a linear combination of states with different numbers of solitons.

\begin{center}
\begin{figure}
\includegraphics[width=\columnwidth,pagebox=artbox]{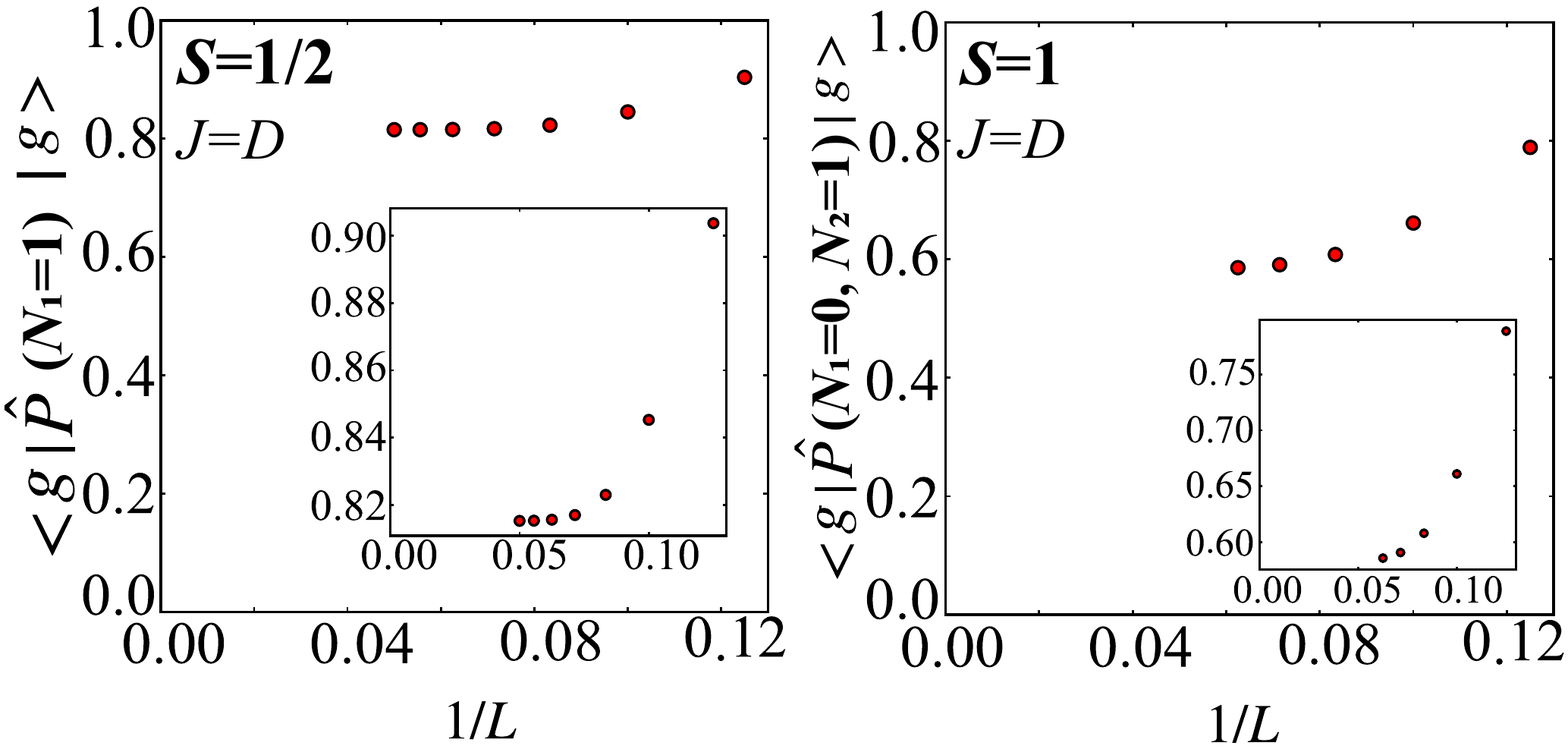}
\caption{Size dependence of the weight of the one-soliton states in the ground state of $\hat{\mathcal{H}}_{\rm ch}$ 
for $J=D$ and $S=1/2$ (left panel) and $S=1$ (right panel). The magnetic field is set at $H/D=0.29$ ($H/D=0.5$) for $S=1/2$ ($S=1$). The insets of both panels magnify the vertical axis.}
\label{fig: overlapS0.5}
\end{figure}
\end{center}

We examine numerically to what extent the single soliton basis (the set of eigenstate of $\hat{N}$ with $N=1$) accounts for the ground state wavefunction of $\hat{\mathcal{H}}_{\rm ch}$. 
We set $J=D$. 
Slightly below the critical field, and for site numbers up to $L=20$ for $S=1/2$ ($L=16$ for $S=1$), we find that 80 (58) percent of the weight of the exact ground state $|{\rm g}\rangle $ of $\hat{\mathcal{H}}_{\rm ch}$ is made up of states belonging to the single soliton basis.
Details of this examination is shown in  
Fig.~\ref{fig: overlapS0.5}, where for the $S=1/2$ case, the weight 
of the one-soliton state within the ground state of $\hat{\mathcal{H}}_{\rm ch}$, i.e. 
$\langle {\rm g}|\hat{P}(N_1=1)|{\rm g}\rangle$ 
is displayed as a function of $1/L$ in the left panel. 
Likewise, for $S=1$ (right panel), we have evaluated the corresponding weight $\langle {\rm g}|\hat{P}(N_1=0,N_2=1)|{\rm g}\rangle$. 
We take these results to be a strong indication that the physical picture derived rigorously in the $J=0$ limit continues to be valid qualitatively even when $J$ is comparable to $D$. 
\section{Discussions}\label{sec: discussion}
In the preceding, we established 
that in the p$DH$ models for  
arbitrary $S$, a soliton whose height is 
$f$  
has a crystal momentum of $\pi f$ in its lowest energy state.  
More generally, we verified in a rigorous manner 
that the lowest energy state belonging to 
the sector $(N_1, \cdots, N_{2S})$, where 
$N_f (1\le f \le 2S)$ is the number of 
solitons of amplitude $f$ present in this state, has the crystal momentum $k=\pi\sum_f f N_f$. 
We have termed our finding the 
{\it height parity effect}. 
To demonstrate the power of this result, 
we note that it immediately rules out 
a spin parity effect for a {\it spin wave}, which is identified 
with a soliton of height $f=1$ and length $1$: 
the crystal momentum of this state is $k=\pi$  
irrespective of the value of $S$. 

Numerical calculations  
show that only solitons with the  
maximal height $f=2S$ contribute to the ground state of the p$DH$ model.
This implies that the height parity effect for the ground state of this model is  
 $k=2\pi S N_{2S}$, which is precisely the observed spin parity effect.

The magnetization processes of 
the p$DH$ models for each $S$ 
consist of successive level crossings from a $N_{2S}$-soliton state to a $N_{2S}-1$-soliton state. For half-integer $S$, each level crossing is protected by the soliton number (which changes by $\Delta N_{2S}=-1$) and the crystal momentum (which changes by $\Delta k=\pi$) while 
it is protected only by the soliton number for integer $S$. Switching on the exchange interaction, the soliton numbers are no longer conserved quantities and thus the level crossing for integer $S$ turns into a crossover, while it remains  
protected by a 
$\pi$-shift in the momentum for half-integer $S$ up to a certain magnitude of exchange interaction $J$. Although this threshold value of $J$ is unknown, our numerical results for magnetization curves and the relative weight of one-soliton state in the ground state for $S=1/2$ and $S=1$ implies that the spin parity effect in monoaxial chiral magnets with $J/D\leq 1$ is well captured by the properties of quantum solitons developed in the present study. 

We should caution the reader, though that 
determining whether a spin parity effect is 
present as well in monoaxial chiral magnets  %
with a 
small $S$ and $J\gg D$ %
as is typical in existing magnets, 
will require further investigations. 
If such an effect is indeed verified in systems 
belonging to the ^^ ^^ solid state" limit 
$J\gg D$, it is premature with the 
information at hand to claim that 
its underlying mechanism, 
along with the proper characterization and definition of a quantum soliton, 
is the same as those described in this paper, 
which are valid 
under the condition $J\le D$. 

That said, it is nevertheless interesting to compare notes with 
the semiclassical approach, which is considered to work at large $S$ and in the wide-soliton regime $J>D$. This method is described in some detail in Appendix A. The salient points are (1) the recovery of the spin parity effect $k=2\pi S$, and (2) a gauge structure inherent to the effective action which can be viewed as roughly corresponding to the signed basis argument of the main text. While these analogies certainly appear to point to universal aspects which arise for deeper topological reasons that 
hold irrespective of the specific regime of interest, a firmer 
understanding on this point remains to be established. 

Experimental realizations of the limit $D\gg J$ in 
non-solid state settings is another 
direction worth pursuing. 
In particular, 
the $S=1/2$ $DH$ model,  
if realized 
will 
serve as  
an ideal platform  
for studying the quantum dynamics of solitons.   
On this front we mention that a proposal has recently been made to realize a system equivalent to %
this model using Rydberg atoms
\cite{kunimi}.

Among other apparently significant issues that remain, is a thorough investigation into possible spin parity effects for general $S$ from a purely quantum approach for the following systems: 
antiferromagnetic chiral magnets in 1d\cite{Hongo2020}  and 2d, 2d chiral ferromagnets accommodating skyrmions \cite{Takashima2016,OchoaTserkovnyak2019},
and non-chiral magnets 
with stable solitons arising from an 
Ising anisotropy\cite{BraunLoss1996}. We hope that the 
present work will inspire  
activities in this direction 
that will go a long way 
toward painting a coherent picture for 
spin parity effects in quantum magnets.
\section{Summary}\label{sec: conclusion}
In summary, we have numerically verified and subsequently tracked down the mechanism responsible for a spin parity effect which is present in the ground state of a monoaxial chiral ferromagnet spin chain. 

Our study started with a numerical evaluation of the magnetization curve for finite sized systems falling within the 
regime $J\le D$. For half-odd integer $S$, the curve consists of level crossings accompanied by a jump of the crystal momentum by the amount $\pi$. The behavior is very different when $S$ is integral: 
the curve is continuous, features crossover events, and the ground state's crystal momentum remains zero throughout. 

To get a handle on this problem, 
we constructed a limiting-case Hamiltonian, the $S=1/2$ $DH$ model, 
where the soliton number is a conserved quantity. We established rigorously that the lowest-energy state with $N$ solitons has the crystal momentum $\pi N$. 

Encouraged by this result, 
we constructed a natural generalization 
of the $DH$ model to arbitrary $S$, 
the p$DH$ model 
(which reduces to the former when $S=1/2$).  Quantum solitons in the $S_z$ basis, of integer-valued heights ranging from 1 to $2S$, are all conserved quantities of this model. Let $N_f$ be the numbers of height-$f$ solitons that are 
present in a given state. We showed rigorously that the lowest energy state 
within the sector $(N_1 , \cdots, N_{2S})$ possesses the crystal momentum 
$\pi \sum_{f} f N_f$ (the height parity effect). The spin parity effect
$k=2\pi S N_{2S}$, which is realized in the ground state of this model follows from the height parity effect when all $N_{f}$s are zero  with the sole exception of $N_{2S}$.

The p$DH$ model thus allows for an 
interpretation of the spin parity effect 
which governs its magnetization process in terms of sharply-defined quantum solitons. It also serves to provide a physical picture for the same effect which is observed in the more general model of a monoaxial chiral ferromagnet with a finite $J$ when $J/D \leq 1$. We have confirmed numerically that the picture derived from the p$DH$ model holds up in this regime. 
\begin{acknowledgments}
This work was supported by JSPS KAKENHI Grants Number 20K03855, 21H01032 (YK) and 19K03662 (AT). We thank H. Katsura for  
informative discussions, and Y. Suzuki and K. Omiya for their helpful suggestions on the proof concerning the p$DH$ model. We also thank J. Kishine, Y.~Togawa, S.~C.~Furuya, M.~Kunimi, and T.~Tomita for sharing with us their valuable insights 
on monoaxial chiral magnets. The computation in this work 
was performed 
using the facilities of the Supercomputer Center, the Institute for Solid State Physics, the University of Tokyo. 
Numerical calculations were performed using the $\mathcal{H}\Phi$
package\cite{HPhi}.
\end{acknowledgments}
\appendix
\section{The semiclassical approach }
In this appendix 
we record for completeness what a semiclassical treatment using the spin coherent state path integral, which is valid at large $S$ and under the condition $J>D$, says about the quantum mechanical features of solitons that we have discussed in the main text. 
Much of what follows borrows heavily from the work of Braun and Loss\cite{BraunLoss1996} on the quantum dynamics of solitons in 
{\it nonchiral}  
magnets. Differences that arise in the chiral counterpart will be highlighted as they appear. As mentioned in the main text (see the Discussion section), 
it must be stressed that extrapolating the results of a semiclassical analysis to 
the regime relevant to the present paper is not straightforward. Still the reader will notice  interesting parallels between the two approaches, which we believe is well worth appreciating.

\subsection{Long wavelength effective action}
We take up the same Hamiltonian as in the main text:

\begin{eqnarray}
{\cal H}&=&-J\sum_{\langle ij \rangle}{\bm S}_{i}\cdot{\bm S}_{j}
-h\sum_{j} 
S_{j}^{z}
 + K\sum_{j} 
\left(
S_{j}^y
\right)^2
\nonumber\\
&&-\sum_{\langle ij \rangle} 
D{\hat y}
\cdot {\bm S}_{i} \times {\bm S}_{j}.
\end{eqnarray}

\noindent 
where $J, K>0$. In this appendix 
we will choose to align the spin chain with the $y$-axis. 

In the following we will work in Euclidean space-time and put $\hbar=1$. 
The spin coherent state path integral approach then consists of 
writing each spin vector as c-numbered entities ${\bm S}_{j}=S{\bm n}_{j}$ 
where ${\bm n}_{j}^2 =1$, and studying the action 
\begin{equation}
{\cal S}[\{ {\bm n}_{j}(\tau, x)\}]=
{\cal S}_{\rm BP}+\int d\tau {\cal H}.
\end{equation}
The first term ${\cal S}_{\rm BP}$ records the spin Berry phase, i.e.  
\begin{eqnarray}
{\cal S}_{\rm BP}&=&\sum_j iS\omega[{\bm n}_{j}(\tau)]
\nonumber\\
&=&
\sum_{j}iS(1-\cos\theta_j (\tau))\partial_{\tau}\phi_{j}(\tau),
\label{BP action}
\end{eqnarray}
where $\omega[{\bm n}_{j}(\tau)]$ is the solid angle traced out on the 
unit sphere by the vector ${\bm n}_j (\tau)$ in the course of its 
imaginary-time evolution. In the second line  
we have introduced the spherical coordinates $(\theta, \phi)$ via 
\begin{equation*}
n_x = \sin\theta\sin\phi, n_y =\cos\theta, n_z = \sin\theta\cos\phi. 
\end{equation*} 

\noindent
Below we take the continuum limit and seek the low energy effective action 
for our system.  
Assume that $K\gg J$. 
Since we then expect $n_y (\tau,y)$,  the hard-axis component of ${\bm n}$ 
to be sufficiently small 
compared to the portion lying within the easy ($zx$-)plane, 
it suffices to employ the parametrization $\theta=\pi/2-\delta\theta$, 
and ignore the periodic nature of the angular variable $\delta\theta$. The $2\pi$-periodicity of $\phi$, on the other hand is essential for keeping track 
of solitons and will be retained. 
We now expand each term in the Hamiltonian ${\cal H}$ up to quadratic order 
in $\delta \theta$. 
\begin{figure}[h]
\begin{center}
\includegraphics
[height =8cm, bb=0 0 1300 620]
{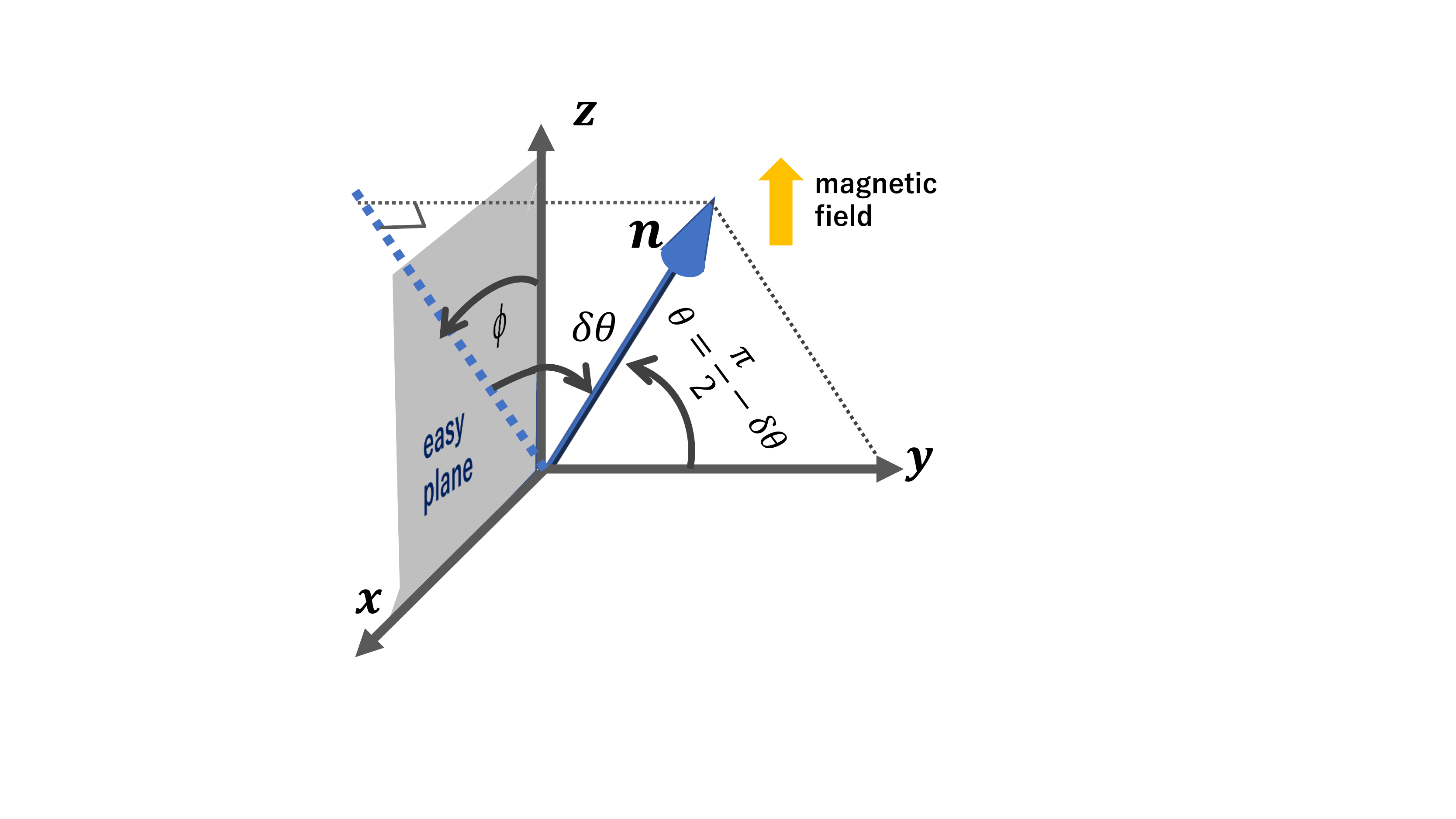}
\end{center}
\vspace{-15mm}
\caption{Spherical coordinate and geometry used in the text.} 
\label{sphericalcoordinate}
\end{figure}
Collecting $\delta \theta$-related contributions, we have 
\begin{equation*}
{\cal L}_{\delta\theta}
=
\frac{KS^2}{a}(\delta\theta)^2 +\frac{JS^2}{2a}(\partial_y \delta\theta)^2 -i\frac{S}{a}(\partial_{\tau}\phi)\delta\theta.
\end{equation*}
The notation $a$ stands for the lattice constant. 
The last term on the right comes from the Berry phase action ${\cal S}_{\rm BP}$. 
Focusing on the long wavelength regime where only  
Fourier modes satisfying $k_y\ll 1\ll \sqrt{\frac{K}{J}}$ are incorporated, 
we can readily integrate over the $\delta\theta$ fluctuations in ${\cal L}_{\delta \theta}$, 
which leaves us with a new term 
$\frac{S}{4Ka}(\partial_{\tau}\phi)^2$.  (Physically this can be understood by 
noting that since, from (\ref{BP action}), $n_{y}$ is canonically conjugate to $\phi$, 
the hard axis anisotropy term $\sim n_{y}^2$ can be traded for the kinetic energy 
related to the dynamics of $\phi$, i.e. $\sim {\dot{\phi}}^2$. )
On combining with the 
remaining terms, we arrive at 
an effective action ${\cal S}_{\rm eff}[\phi(\tau,x)]=\int d\tau dx {\cal L}_{\rm eff}$, 
where 
\begin{eqnarray}
{\cal L}_{\rm eff}&=&
i\frac{S}{a}\partial_{\tau}\phi+\frac{JS^2}{2a}\left[\frac{1}{{c_s}^2}(\partial_{\tau}\phi)^2 +(\partial_y \phi)^2 \right] \nonumber\\ 
&&-\frac{D S^2}{a}\partial_y \phi 
-\frac{hS}{a^2}\cos\phi,
\label{effective action1}
\end{eqnarray}
and $c_{s}=\sqrt{KJSa}$. 
\noindent 
This is a chiral variant of the quantum sine-Gordon action. 
As we will shortly see, the first entry on the right hand side, 
descending from ${\cal S}_{\rm BP}$,  
is a {\it topological term} which is ultimately responsible for the 
occurrence of the spin parity effect of soliton excitations 
as viewed in this semiclassical language. 

\subsection{Soliton collective coordinates}

We proceed to extract from (\ref{effective action1}) 
information pertaining to the   
soliton dynamics.  We will focus for simplicity on 
the one-soliton sector, though similar analysis carries through  
for more complex situations. 
We begin by observing that the 
Euler-Lagrange equation which follows from ${\cal S}_{\rm eff}$ 
is the quantum sine-Gordon equation 
\begin{equation}
\frac{1}{c_s^2}\phi_{\tau\tau}-\phi_{yy}=M^2 \sin\phi, \hspace*{3mm} M^2 =\frac{h}{JSa}, 
\label{qSG eq}
\end{equation}
where subscripts stand for derivatives.  
Static soliton and antisoliton solutions of height $2\pi$ can be written explicitly as  
\begin{equation}
\phi_0 (y)=\pm4\tan^{-1} e^{M(y-Y)},
\label{SG soliton configs}
\end{equation}
as can easily be verified by direct inspection. The letter $Y$ in the above stands for 
the center coordinate of the soliton.  
While the DM term, being a total derivative, does not enter into the equation of motion (\ref{qSG eq}), it plays the important role of 
selecting out the energetically favorable sign in (\ref{SG soliton configs}), 
which in the present convention is positive. 
(One needs to incorporate it explicitly though when incommensurability effects set in at lower magnetic fields.) 

We are now ready to promote the soliton's position $Y$ to 
a dynamical collective coordinate $Y(\tau)$. 
Plugging the configuration 
$\phi_0 (y-Y(\tau))$ into (\ref{effective action1}), we find 
\begin{equation}
{\cal S}_{Y}[Y(\tau)]=\int d\tau \left[
-i\frac{2\pi S}{a}\dot{Y}+\frac{1}{2\tilde M}\dot{Y}^2
\right],
\end{equation} 
where ${\tilde M}=\frac{a}{8JS^2 M}$. 
Here we have assumed a 
soliton of height $2\pi$, or equivalently a configuration for which the winding number  
$Q\equiv\frac{1}{2\pi}\int_{-\infty}^{\infty}dy \partial_y \phi_0$ is unity. 
The generalization to arbitrary $Q$ is straightforward, 
where, in particular the first term on the right is simply multiplied by that integer.  
This is formally equivalent to 
the action of a charged point particle, with the first term 
representing the coupling between the particle and a (Berry) gauge field. 
We will see in a moment that this coupling term, despite it being a total derivative, 
is absolutely crucial for arriving at the correct quantum mechanical features.     

It is natural to expect that  
the soliton, now viewed as a quantum mechanical particle 
hopping through the spin chain, also experiences an 
effective periodic potential $V(y)$ with the property $V(y+a)=V(y)$, 
which imprints 
the underlying lattice structure on the dynamics and thus 
renders the soliton to form Bloch bands.  
We refer the reader to Braun and Loss\cite{BraunLoss1996} for an explicit 
evaluation of this potential, which can be carried out e.g. by treating 
 inter-site tunneling events in a instanton gas approximation.   
The corresponding Hamiltonian which takes these three terms into account is  
\begin{equation}
{\cal H}=\frac{1}{\tilde M}({\hat P}_{Y}-\frac{2\pi S}{a})^2 +V(Y),
\label{Hamiltonian for Y}
\end{equation}
with ${\hat P}_{Y}$ the momentum operator conjugate to $Y$. 
It is clear from this form that the lowest energy state in the 1-soliton sector 
carries a crystal momentum of 
$p_{\rm sol}=\frac{2\pi S}{a}$, reproducing the findings of the main text. 

Having seen how the spin Berry phase has made its way into the expression for $p_{\rm sol}$, 
it is instructive to perform at this point a 
sanity check: recall that for the Berry phase action (\ref{BP action}) 
we made use of the so-called north-pole gauge, in which the Dirac string goes through the south pole. 
We could have equally well opted to use the south-pole gauge, where 
${\cal S}_{\rm BP}=\sum_{j}iS(-1-\cos\theta_j (\tau))\partial_{\tau}\phi_{j}(\tau)$. 
Repeating the whole procedure for the latter, we find that the value of $p_{\rm sol}$
merely shifts by $\frac{2\pi}{a}$, thereby demonstrating the gauge independence 
of the result.   
The lesson to be learned then, as emphasized 
early on by Haldane\cite{HaldanePRL1986}, 
is that retaining the full expression for the solid angle $\omega$ (whatever choice of gauge one makes) 
is crucial for safely extracting information on the crystal momenta of a ferromagnet.  

As a final note before proceeding, we remark that we 
could have foreseen the value of the crystal momentum determined in this subsection, 
once we have chosen to focus on a 
$2\pi$ soliton: we adopt for this 
purpose 
Haldane's semiclassical theory\cite{HaldanePRL1986} mentioned above,
which states that if a snapshot configuration ${\bm n}(y)$ of a 
1d ferrromagnet obeying a periodic 
boundary condition subtends a solid angle $\omega[{\bm n}(y)]$, 
that state carries the crystal momentum $\frac{S}{a}\omega$. As $\omega=2\pi$ for a $2\pi$ soliton, this relation  reproduces the result 
$p_{\rm sol}=\frac{2\pi S}{a}$. 
(We should also mention that the identification of 
the crystal momentum of a soliton in its lowest energy state, 
with the Berry phase associated with the 
snapshot spin configuration appears in several of the 
earlier work on chiral ferromagnetic spin chains\cite{kishineovchinnikov2015,BostremGalileianSymmetry2008,kishineCoherentSlidingDynamics2012}. 
Implications to spin parity effects or to the magnetization process, however, 
are not considered there.)
\subsection{Effect of magnetic fluctuations}
The foregoing basically followed from a treatment at the saddle point level, 
and as such needs to be 
submitted to a stability analysis against quantum fluctuations, 
i.e. 
the effects of spin wave fluctuations $\varphi$ 
around the moving rigid-soliton configuration $\phi_0$ :   
\begin{equation}
\phi(\tau,y)=\phi_0 (y-Y(\tau))+\varphi(y-Y(\tau),\tau). 
\end{equation} 
As this is not directly related to the spin parity effect, 
we once again refer the interested reader to Braun and Loss\cite{BraunLoss1996,BraunProceedings1995}
for the relevant technical details, and merely state 
the outcomes of this analysis. (An alternative method based on Dirac's formalism 
for constrained quantum theory can be found in the review article of 
Kishine and Ovchinikov\cite{kishineovchinnikov2015}.)
An expansion to second order in the spin wave fluctuation 
yields the spin wave dispersion 
\begin{equation}
\epsilon_{k}=\frac{JS^2}{2a}(k^2 +M^2),
\end{equation}
where one sees that a mass has been induced 
by the Zeeman field. 
Meanwhile a coupling between soliton coordinate $Y$ and the spin wave $\varphi$ enters the action 
at the same order, which can lead to 
damping (memory) effects as well as a renormalization of the soliton's rest mass.  The former is found to have 
a characteristic decay time $\tau=\frac{1}{2{\sqrt{Kh}}}$ which, if sufficiently smaller than the time scale on which 
${\dot Y}$ changes, is negligible. The latter is of the order of ${\cal O}(1/S)$, which in the semi-classical regime 
should also be small.  
\subsection{Implications}
A remark on the behavior of the magnetization curve in light of the 
semiclassical 
effective theory is in order. 
The Hamiltonian (\ref{Hamiltonian for Y}) implies that 
the introduction of an additional $2\pi$ kink into the system, 
i.e. a process for which $\delta Q=1$, is accompanied by 
a change in the crystal momentum by the amount 
\begin{equation*}
\delta P_{Y}=\frac{2\pi S}{a}\delta Q =2\pi S. 
\end{equation*}
Thus for half-integer $S$, momentum conservation prohibits 
tunneling between configurations differing in $Q$ by one. This will 
result in level crossing. In contrast to this, tunneling and hence 
level repulsion can occur when $S$ is integral.  The same 
conclusion also follows from a path integral point of view. 
The kink insertion is a singular space-time process (a phase-slip). 
Consider two different space-time patterns in which such 
events occur, the second one centered at a plaquette (in the $y$ vs $\tau$ plane) immediately to the right
of the first. These two events each enter the path integral with Feynman weights $e^{-{\cal S}_{\rm Event1}}$ and 
$e^{-{\cal S}_{\rm Event2} }$,  
differing only by the phase  $e^{-i2\pi S}$,  
and thereby canceling out when $S$ is half-integral.  Since such pair-wise cancellation occurs generically, 
one concludes that phase slips do not contribute to the partition function.  

The spin Berry phase's influence on soliton 
dynamics can be modulated by the addition of a longitudinal component (i.e. along the chain) to the external 
magnetic field\cite{BraunLoss1996,kishineCoherentSlidingDynamics2012}. 
For the problem at hand, this will have the effect of continuously changing the crystal momentum in proportion to the superimposed field.
At special values of the latter (ideally there are $2S+1$ such values), the magnetization process in the half-odd integer $S$ case (as a function of the transverse field) is expected to mimic the behavior of an integer $S$ spin chain in the absence of the longitudinal field. The manner in which the spin parity effect is in principle 
controllable using an external parameter is reminiscent of what happens in antiferromagnetic chains, when one introduces (and continuously varies the strength of) 
a bond-alternating component to the nearest neighbor exchange interaction
\cite{HaldaneThetaPhysics,HaldaneAffleck1987}.

Finally we recall that recognizing the global structure inherent to the ground state wavefunction  
was the key that lead us to some of the central conclusions of the main text. 
This prompts us to briefly recapitulate the discussions of the preceding paragraphs  
from the vantage point of wave function properties. 
To this end we note that it is possible to envisage a continuum counterpart
\cite{TanakaTotsukaHu2009}
for the 
 ^^ ^^ signed basis'' expansion of the ground state's state vector that was discussed in the main text:
\begin{eqnarray}
|\Psi\rangle&=&\int {\cal D}\phi(y)
e^{-iS\int \frac{dy}{a}\phi(y)} |\phi(y)\rangle\langle\phi(y)|\Psi\rangle
\nonumber\\
&\equiv&
\int{\cal D}\phi(y)
e^{-iS\int \frac{dy}{a}\phi(y)}\Psi[\phi(y)] |\phi(y)\rangle.
\label{continuum state vector}
\end{eqnarray} 
The phase factor $e^{-iS\int \frac{dy}{a}\phi(y)}$ is the continuum analog of 
the all-important kink-counting sign factor $(-1)^{2S \sum_{\rm kink} j_{\rm kink}}$, where 
$j_{\rm kink}$ is the lattice site at the left end of a soliton. (The analogy becomes more transparent upon rewriting this factor as $e^{iS\int \frac{dy}{a}y\phi'(y)}$.) Meanwhile the wave functional 
$\Psi[\phi(y)]$ corresponds to the positive-sign expansion coefficients, and should 
have the property that it can be choosen to be real and nodeless, and exhibit the 
lattice periodicity $\Psi[\phi(y+a)]=\Psi[\phi(y)]$. The crystal momentum 
associated with this state can then be obtained as follows. Writing the generator of 
a one site translation as ${\hat T}$, and in addition defining 
\begin{equation}
{\tilde \Psi}[\phi(y)]\equiv 
e^{-iS\int \frac{dy}{a}\phi(y)}\Psi[\phi(y)],
\end{equation}
we have
\begin{eqnarray*}
{\hat T}{\tilde \Psi}[\phi(y)]&=&{\tilde \Psi}[\phi(y-a)]
\nonumber\\
&=&{\tilde \Psi}[\phi(y)]\times e^{iS\int \frac{dy}{a}
\{
\phi(y)-\phi(y-a)
\}
}
\nonumber\\
&\simeq&
{\tilde \Psi}[\phi(y)]\times
e^{iS\int dy (\partial_y \phi)}
\nonumber\\
&=&
{\tilde \Psi}[\phi(y)]\times
e^{i2\pi S N} {\hspace*{3mm}} N: {\rm soliton}{\hspace*{1.5mm}}{\rm number}.
\end{eqnarray*}
The resemblance with how the lattice wave function transforms under translation is apparent. 
To see how this relates to the semiclassical theory  (\ref{effective action1}), 
we first write down its Hamiltonian {\it for the case where the topological term is absent}. This reads 
\begin{eqnarray}
{\cal H}_{\rm eff}&=&\int dy 
\left[
\frac{a^2 K}{S}
{\hat \pi}_{\phi}^2 
+\frac{JS^2}{2a}(\partial_y \phi)^2 
\right.
\nonumber\\
&&\left. 
{\hspace*{8mm}}
-\frac{D S^2}{a}\partial_y \phi
-\frac{hS}{a^2}\cos\phi
\right].
\end{eqnarray}
Here we have used the notation ${\hat \pi}_\phi(y) \equiv -i\frac{\delta}{\delta \phi(y)}$.  
Let us call the ground state wave functional for this Hamiltonian $\Psi[\phi(x)]$. 
As there are no topological terms which act on the solitons as Aharonov-Bohm like fluxes, 
we expect that $\Psi[\phi(x)]$ can to chosen to be real, is nodeless, and respects the lattice translation symmetry\cite{ZhangZimanSchulz1989}. 
Upon reintroducing the topological term, the momentum ${\hat \pi}_{\phi}$ entering the 
first term on the right hand side of the above equation receives the shift 
${\hat \pi}_{\phi}\rightarrow{\hat \pi}_{\phi}+\frac{S}{a}$. Since this shift is of the form of a 
coupling of a charged matter to a gauge field, it is straightforward to see that the wave functional accordingly 
^^ ^^ gauge transforms" into $e^{-iS\int \frac{dy}{a}\phi(y)}\Psi[\phi(y)]$. The same conclusion is reached by formally expressing the ground state wave functional  as a constrained path integral, i.e.$\Psi[\phi(y)]\propto\int{\cal D}\phi(\tau,y)e^{-{\cal S}_{\rm eff}[\phi]}$, 
where one takes the sum over paths in Euclidean space-time such that the  configuration at the terminal imaginary time (which is taken to be sufficiently large) always ends up as $\phi[y]$. 
In this approach the phase factor $e^{-iS\int \frac{dy}{a}\phi(y)}$ derives from a boundary contribution of the topological term which is generated at the end of the imaginary time axis. (This method is valid provided there is an energy gap between a unique ground state and the excited states.)
\cite{XuSenthil2013,TakayoshiTotsukaTanaka2015}. 

To seek the counterpart of the height parity effect and the $DH$ and p$DH$ models in the semiclassical/field-theoretical framework, as well as to undertake a quest for spin parity effects at much lower magnetic fields 
where incommensurability effects need to be incorporated, are interesting problems that we leave for the future.
%
%
%
%
\clearpage
\renewcommand{\thesubsection}{S\arabic{subsection}}
\renewcommand{\theequation}{S\arabic{equation}}
\setcounter{equation}{0}
\setcounter{subsection}{0}
\renewcommand{\thefigure}{S\arabic{figure}}
\setcounter{figure}{0}
\renewcommand{\thetable}{S\arabic{table}}
\setcounter{table}{0}
\makeatletter
\c@secnumdepth = 2
\makeatother
\onecolumngrid
\begin{center}
 {\large \textmd{Supplemental Materials:} \\[0.3em]
 {\bfseries Spin parity effects in monoaxial chiral ferromagnetic chain }\\
 \ \\
 \textmd{Sohei Kodama, Akihiro Tanaka, and Yusuke Kato}} \\[0.3em]
\end{center}
\setcounter{page}{1}
\subsection{Proof of Lemma 2}
We denote the relation between $\bm{n}$ and $\bm{n}'$ that belong to $V_N$ with $N\in [1,L/2-1]$ by $\bm{n}\sim \bm{n}'$ when 
there exists a positive integer $l$ such that
\begin{equation}
M(\bm{n},\bm{n}';l)>0,\label{eq: Mnnl}    
\end{equation}
with
\begin{equation}
M(\bm{n},\bm{n}';l):=(-1)^{\delta(\bm{n})+\delta(\bm{n}')}\langle\bm{n}|(-\hat{\mathcal{H}}_{\rm DM})^l|\bm{n}'\rangle.\label{eq: irreducibility-S-1/2-proof}
\end{equation}
We prove that $\bm{n}\sim \bm{n}'$
 for arbitrary pairs of 
$\bm{n}$ and $\bm{n}'$ belonging to $V_N$ with $N\in [1,L/2-1]$. The off-diagonal matrix elements of $\hat{\mathcal{H}}_{DH}$ in this basis stem from $\hat{\mathcal{H}}_{\rm DM}$. Let the minimum $l$ satisfying Eq.~\eqref{eq: Mnnl} be $l_0$. Then  
\begin{equation}
(-1)^{\delta(\bm{n})+\delta(\bm{n}')}\langle\bm{n}|(-\hat{\mathcal{H}}_{DH})^{l_0}|\bm{n}'\rangle=M(\bm{n},\bm{n}';l_0)>0     
\end{equation}
follows. \ \\
{\bf Lemma S1}\\
The relation $\sim$ is transitive and symmetric.
\begin{proof}
(symmetric) $\bm{n}\sim \bm{n}'\leftrightarrow \bm{n}'\sim \bm{n}$ because matrix elements of $\hat{\mathcal{H}}_{\rm DM}$ in the present basis is real. \\  
(transitive) When $\bm{n}_{\rm I} \sim \bm{n}_{\rm II}$ and $\bm{n}_{\rm II} \sim \bm{n}_{\rm III}$ for $\bm{n}_{\rm I},\bm{n}_{\rm II},\bm{n}_{\rm III}\in V_N$ with $N\in [1,L/2-1]$, there exist positive integers $l_1$ and $l_2$ such that 
\begin{equation}
M(\bm{n}_{\rm I},\bm{n}_{\rm II};l_1)>0, \quad   
M(\bm{n}_{\rm II},\bm{n}_{\rm III};l_2)>0.    
\end{equation}
It then follows that 
\begin{align}
&M(\bm{n}_{\rm I},\bm{n}_{\rm III};l_1+l_2)\nonumber\\
=&\sum_{\bm{n}'\in V_N}M(\bm{n}_{\rm I},\bm{n}';l_1)M(\bm{n}',\bm{n}_{\rm III};l_2)\\
=&\underbrace{M(\bm{n}_{\rm I},\bm{n}_{\rm II};l_1)}_{>0}\underbrace{M(\bm{n}_{\rm II},\bm{n}_{\rm III};l_2)}_{>0}\nonumber\\
&+\sum_{\bm{n}'(\ne \bm{n}_{\rm II})\in V_N}\underbrace{M(\bm{n}_{\rm I},\bm{n}';l_1)}_{\ge 0}\underbrace{M(\bm{n}',\bm{n}_{\rm III};l_2)}_{\ge 0}\\
&>0.
\end{align}
\end{proof}

For a given $\bm{n}$, let $B=B(\bm{n})$ be the set of $i$ such that $n_{i-1}\ne n_i$.
We label these elements $B(\bm{n})=\{b_\alpha\}$
in the increasing order
\begin{align}
1\le b_1<b_2\cdots<b_{2N}\le L.
\end{align}
We define $m_\alpha$=0 or 1 for $\alpha=[1,2N]$ by 
\begin{equation}
m_\alpha=n_i,\mbox{ with }i=b_\alpha.    
\end{equation}
We also define $b_{2N +1}$ and $m_{2N +1}$ as $b_1 +L$ and $m_1$. 
By definition, $m_\alpha\ne m_{\alpha+1}$ for  $\alpha=[1,2N]$.   

\begin{center}
\begin{figure}
\includegraphics[width=0.4\columnwidth,pagebox=artbox]{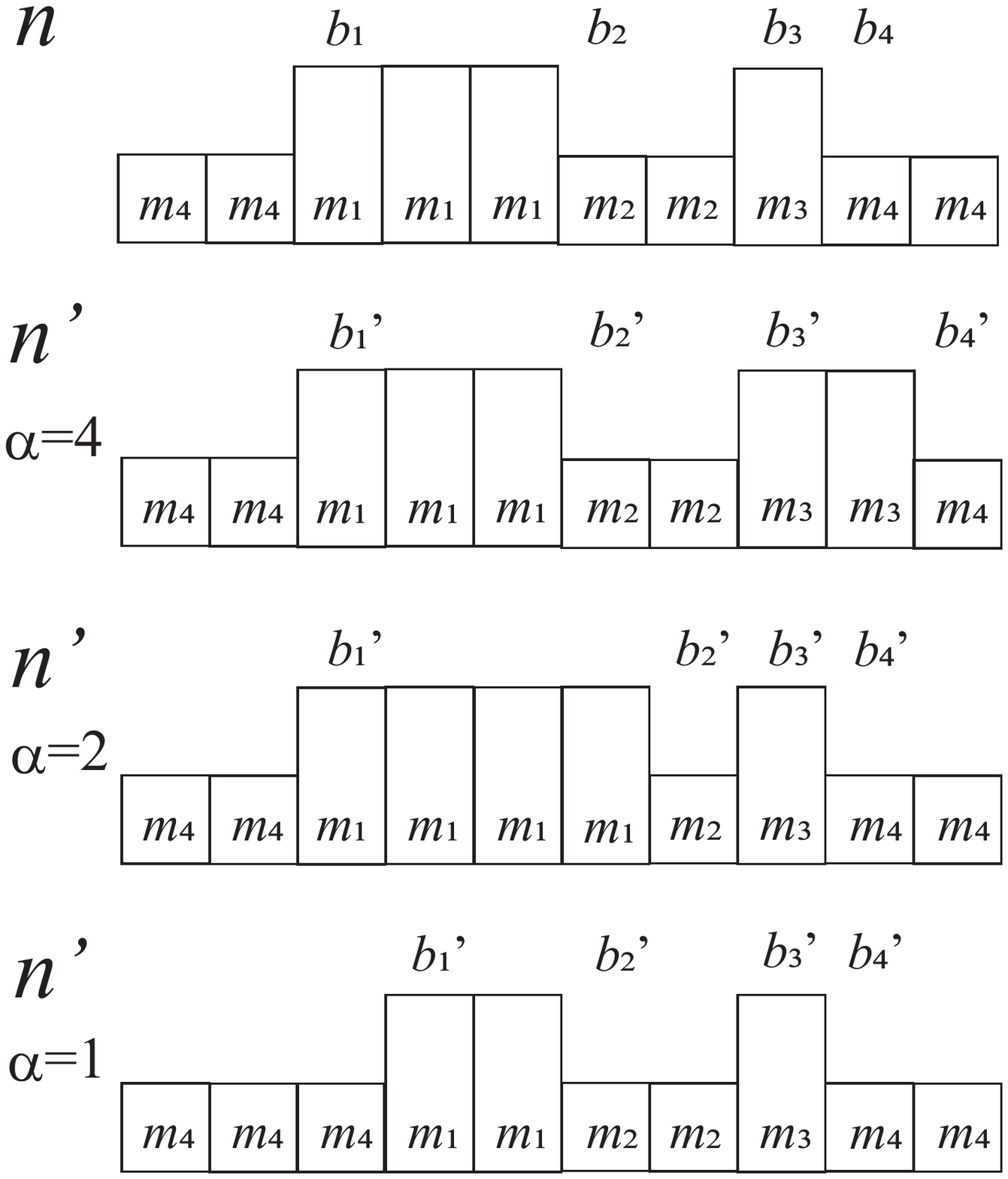}
\caption{Upper-most figure schematically shows an example Eq.~\eqref{eq: example-of-balpha-malpha} of $\{b_\alpha,m_\alpha\}_{\alpha=1}^{2N}$ for $N=2$ and $L=10$. Other figures show examples of $\bm{n}'$ satisfying Eq.~\eqref{eq: b-nprime-n-relation-S=1/2}.}
\label{fig: FigS_lemmaS2}
\end{figure}
\end{center}

When $L=10$ and $\bm{n}=0011100100\in V_2$, for example (see the upper-most picture in Fig.~\ref{fig: FigS_lemmaS2}), 
\begin{subequations}
\begin{align}
&b_1=3,\quad b_2=6,\quad b_3=8,\quad b_4=9,\quad b_5(=b_1 +10)=13 \\    
&m_1=1,\quad m_2=0,\quad m_3=1,\quad m_4=0,\quad m_5(=m_1)=1. 
\end{align}
\label{eq: example-of-balpha-malpha}
\end{subequations}
The states $|\bm{n}\rangle\in V_N$ with $N\in [1,L/2-1]$ contain at least a pair of adjacent sites with spin contents $^^ ^^ 11"$ or $^^ ^^ 00"$ (otherwise $\bm{n}$ would belong to $V_0$ or $V_{L/2}$) and thus also a segment of consecutive three sites with spin contents $^^ ^^ 011"$ or $^^ ^^ 100"$.
In our notation, $^\exists\alpha\in [1,2N],\quad b_{\alpha+1}-b_{\alpha}\ge 2$.

The next lemma implies that an action of $\hat{\mathcal{H}}_{\rm p0}$ on $|\bm{n}\rangle$ can shift one of the boundaries $b_\alpha(\bm{n})$ by 
one site to the right when $b_{\alpha+1}(\bm{n})-b_\alpha(\bm{n})\ge 2$. 

\begin{center}
\begin{figure}
\includegraphics[width=0.5\columnwidth,pagebox=artbox]{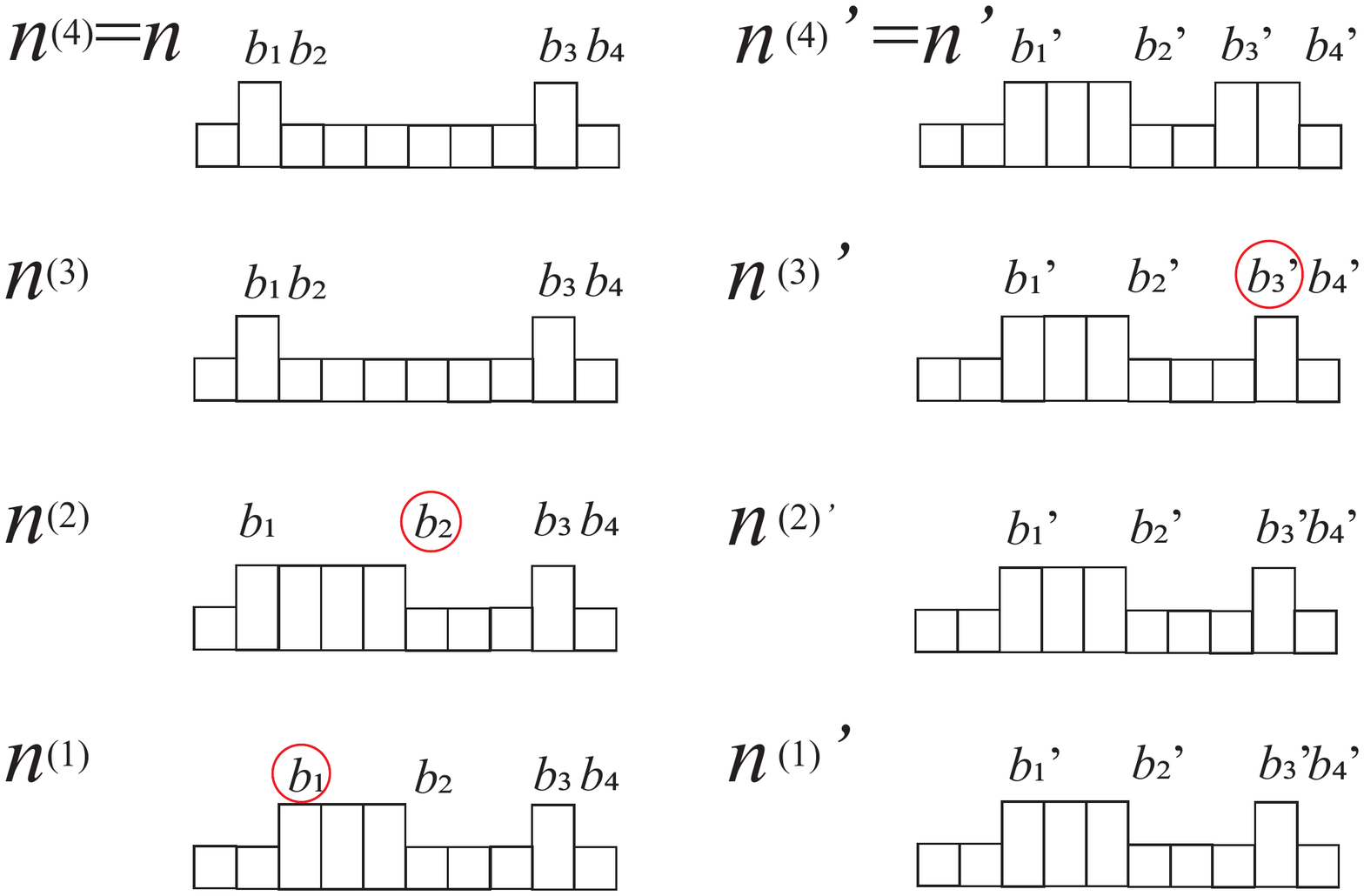}
\caption{Examples of $\{\bm{n}^{(\alpha)},\bm{n}^{'(\alpha)}\}_{\alpha=1}^{2N}$, which are defined, respectively, by Eqs.~\eqref{eq: def-n-alpha} and \eqref{eq: def-n-prime-alpha}. The characters for updated $b_\alpha$s are encircled.}
\label{fig: FigS_n_to_alpha}
\end{figure}
\end{center}

\begin{center}
\begin{figure}
\includegraphics[width=0.3\columnwidth,pagebox=artbox]{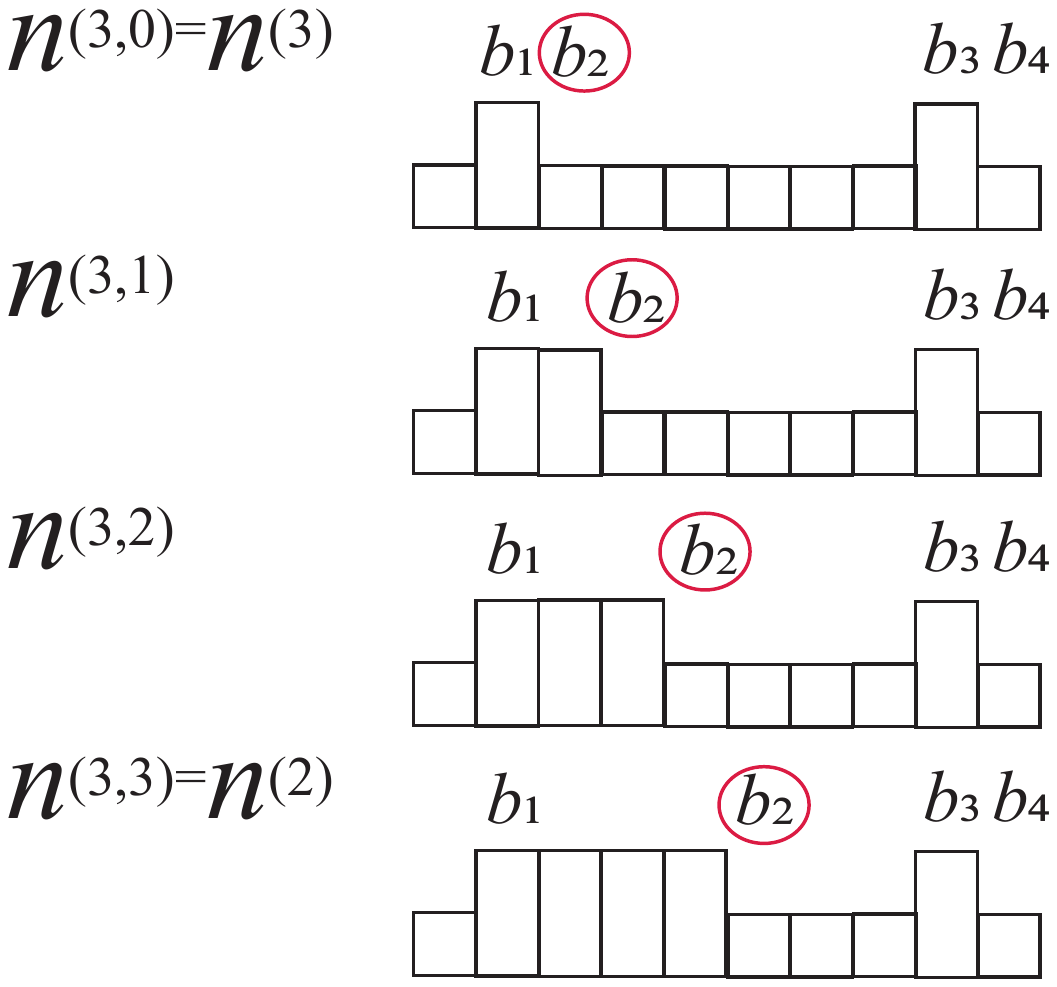}
\caption{Examples of $\bm{n}^{(\alpha,a)}$ defined by Eq.~\eqref{eq: n-alpha-a}. The characters for updated $b_\alpha$s are encircled.}
\label{fig: FigS_n_to_alpha-a}
\end{figure}
\end{center}

{\bf Lemma S2}\\
When $b_{\alpha+1}-b_{\alpha}\ge 2$ for $^\exists\alpha\in [1,2N]$ in $\bm{n}\in V_N$ with $N\in [1,L/2-1]$, $\bm{n}\sim \bm{n}'$, where
\begin{subequations}
\begin{align}
b_\beta(\bm{n}')&=b_\beta(\bm{n})+\delta_{\beta,\alpha}\\
m_\beta(\bm{n}')&=m_\beta(\bm{n})
\end{align}
\label{eq: b-nprime-n-relation-S=1/2}
\end{subequations}
for $\beta\in [1,2N]$. 
Figure \ref{fig: FigS_lemmaS2} shows examples of $\bm{n}$ and $\bm{n}'$ satisfying Eq.~\eqref{eq: b-nprime-n-relation-S=1/2}. 
\begin{proof}
The spin configuration $\bm{n}$ in the three consecutive sites  $i=b_\alpha-1,b_\alpha,b_\alpha+1$ is 
\begin{align}
\bm{n}=(\cdots,m_{\alpha-1},m_\alpha,m_\alpha,\cdots). 
\label{eq: n-three-sites-s=1/2}
\end{align}
It follows from Eq.~\eqref{eq: hi-action-summary-signed basis} that 
\begin{align}
M(\bm{n}',\bm{n};1)
=\frac{D}{2}
(-1)^{\delta(\bm{n})+\delta(\bm{n}')}\langle \bm{n}'|\hat{h}_{b_\alpha}|\bm{n}\rangle>0.  
\end{align}
\end{proof}
{\bf Lemma S3}\\
For $|\bm{n}\rangle\in V_N$ with $N\in [1,L/2-1]$, 
\begin{equation}
T(\bm{n})\sim \bm{n}.\label{eq: Tn-n-transitive-S=1/2}
\end{equation}
\begin{proof}
The relation between $\bm{n}$ and $T(\bm{n})$
is expressed as
\begin{subequations}
\begin{align}
b_\beta(T(\bm{n}))&=b_\beta(\bm{n})+1\\
m_\beta(T(\bm{n}))&=m_\beta(\bm{n})
\end{align}
\label{eq: b-shift-n-and-Tn-S=1/2}
\end{subequations}
for $\beta\in [1,2N]$.     

For $\bm{n}\in V_N$ with $N\in [1,L/2-1]$, 
$^\exists\alpha\in [1,2N],\quad b_{\alpha+1}-b_{\alpha}\ge 2$.

For $\bm{n}'$ defined by Eq.~\eqref{eq: b-nprime-n-relation-S=1/2}, we note that $b_{\alpha}(\bm{n}')-b_{\alpha-1}(\bm{n}')=b_{\alpha}(\bm{n})-b_{\alpha-1}(\bm{n})+1\ge 2$ and thus find that we can shift $b_{\alpha-1}(\bm{n}')$ by 
one site to the right by 
an action of 
$\hat{\mathcal{H}}_{\rm p0}$ on $|\bm{n}'\rangle$,  
\begin{equation}
M(\bm{n}'',\bm{n}';1)
=\frac{D}{2}>0, \label{eq: HM0-ndoubleprime-and-nprime-S=1/2} 
\end{equation}
i.e., $\bm{n}'\sim \bm{n}''$
for $\bm{n}'$ and $\bm{n}''$ defined by
\begin{subequations}
\begin{align}
b_\beta(\bm{n}'')&=b_\beta(\bm{n}')+\delta_{\beta,\alpha-1}\\
m_\beta(\bm{n}'')&=m_\beta(\bm{n}')
\end{align}
\label{eq: b-ndoubleprime-nprime-relation-S=1/2}
\end{subequations}
for $\beta\in [1,2N]$. Note that $b_{\alpha-1}(\bm{n}'')-b_{\alpha-2}(\bm{n}'')=b_{\alpha-1}(\bm{n}')-b_{\alpha-2}(\bm{n}')+1\ge 2$ and thus we can shift $b_{\alpha-2}(\bm{n}'')$ by 
one site to the right in a way similar to the above procedure. By repeating these procedures as schematically shown in Fig.~\ref{fig: FigS_lemmaS3}, we can shift all $b_\beta(\bm{n})$ for 
$\beta\in [1,2N]$ by 
one site to the right and arrive at 
\begin{equation}
\bm{n}\sim \bm{n}'\sim \bm{n}''\sim \cdots \sim T(\bm{n}),     
\end{equation}
i.e., Eq.~\eqref{eq: Tn-n-transitive-S=1/2}.    
\begin{center}
\begin{figure}
\includegraphics[width=0.4\columnwidth,pagebox=artbox]{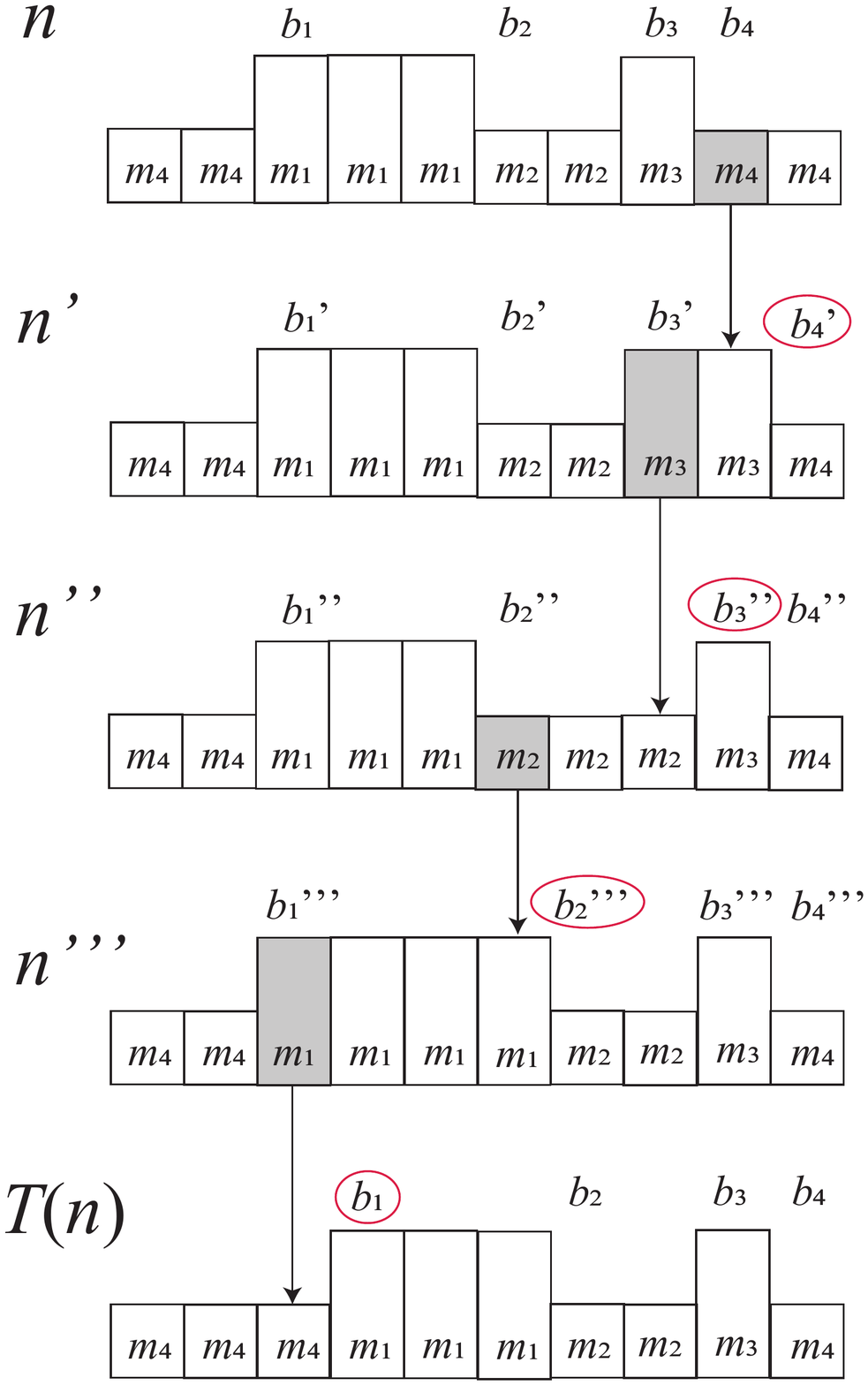}
\caption{Schematics illustrating relation $\bm{n}\sim T(\bm{n})$ (Lemma S3). The arrows represent the action of the local Hamiltonian $\hat{h}_i$, which changes $n_i$ represented by shaded squares by one. The characters for updated $b_\alpha$s by this action are encircled.}
\label{fig: FigS_lemmaS3}
\end{figure}
\end{center}
\end{proof}
{\it Proof of Lemma 2}\\
\begin{proof}
For arbitrary $\bm{n}$ and $\bm{n}' \in V_N$ with $N\in [1,L/2-1]$, there exists an integer $l$ such that 
\begin{subequations}
\begin{align}
b_{2N}(T^l (\bm{n}'))&=b_{2N}(\bm{n})\\
m_\beta(T^l (\bm{n}'))&=m_\beta(\bm{n})
\end{align}
\label{eq: b-Tnprime-n-relation-S=1/2}
\end{subequations}
for $\beta\in [1,2N]$. 
It thus suffices to show that $\bm{n}\sim \bm{n}'$ for $\bm{n},\bm{n}' \in V_N$ with $N\in [1,L/2-1]$ satisfying 
\begin{subequations}
\begin{align}
b_{2N}(\bm{n}')&=b_{2N}(\bm{n})\\
m_\beta(\bm{n}')&=m_\beta(\bm{n})
\end{align}
\label{eq: n-nprime-condition}
\end{subequations}
for $\beta\in [1,2N]$.
For $\bm{n}$ and $\bm{n}'$ satisfying Eq.~\eqref{eq: n-nprime-condition}, we define $\{\bm{n}^{(\alpha)},\bm{n}^{'(\alpha)}\}_{\alpha=1}^{2N}\in V_N$ with $N\in [1,L/2-1]$ by 
\begin{subequations}
\begin{align}
b_{\beta}(\bm{n}^{(\alpha)})&=\left\{
\begin{array}{cl}
\mbox{Max}(b_{\beta}(\bm{n}'),b_{\beta}(\bm{n})),   & \beta\in [\alpha,2N] \\
b_{\beta}(\bm{n}),
     & \mbox{otherwise} 
\end{array}
\right.\\
m_\beta(\bm{n}^{(\alpha)})&=m_\beta(\bm{n}),\quad\beta\in [1,2N]
\end{align}
\label{eq: def-n-alpha}
\end{subequations}
and
\begin{subequations}
\begin{align}
b_{\beta}(\bm{n}^{'(\alpha)})&=\left\{
\begin{array}{cl}
\mbox{Max}(b_{\beta}(\bm{n}'),b_{\beta}(\bm{n}))   & \beta\in [\alpha,2N] \\
b_{\beta}(\bm{n}')
     & \mbox{otherwise}  
\end{array}
\right.\\
m_\beta(\bm{n}^{'(\alpha)})&=m_\beta(\bm{n}'),\quad\beta\in [1,2N].
\end{align}
\label{eq: def-n-prime-alpha}
\end{subequations}
By definition, 
\begin{subequations}
\begin{align}
\bm{n}^{(2N)}&=\bm{n}\\    
\bm{n}^{'(2N)}&=\bm{n}'\\
\bm{n}^{(1)}&=\bm{n}^{'(1)}.
\end{align}
\label{eq: n-alpha-limiting}
\end{subequations}
Examples of $\{\bm{n}^{(\alpha)},\bm{n}^{'(\alpha)}\}_{\alpha=1}^{2N}\in V_N$ for $N=2$ are shown in Fig.~\ref{fig: FigS_n_to_alpha}. 

We show that $\bm{n}^{(\alpha)}\sim \bm{n}^{(\alpha-1)}$ for $\alpha\in [2,2N]$.
When $b_{\alpha-1}(\bm{n})\ge b_{\alpha-1}(\bm{n}')$, $\bm{n}^{(\alpha)}= \bm{n}^{(\alpha-1)}$ and thus we focus on the case where $b_{\alpha-1}(\bm{n})< b_{\alpha-1}(\bm{n}')$. We introduce $\bm{n}^{(\alpha,a)}$ for $a\in [0,b_{\alpha-1}(\bm{n}')-b_{\alpha-1}(\bm{n})]$ by
\begin{subequations}
\begin{align}
b_{\beta}(\bm{n}^{(\alpha,a)})&=\left\{
\begin{array}{cl}
b_{\alpha-1}(\bm{n}^{(\alpha)})+a,
   & \beta=\alpha-1\\
b_{\beta}(\bm{n}^{(\alpha)}),
     & \mbox{otherwise}
\end{array}
\right.\\
m_{\beta}(\bm{n}^{(\alpha,a)})
&=m_\beta(\bm{n}),\quad\beta\in [1,2N].
\end{align}
\label{eq: n-alpha-a}
\end{subequations}
Note that 
\begin{equation}
\bm{n}^{(\alpha,0)}=\bm{n}^{(\alpha)},\quad 
\bm{n}^{(\alpha,b_{\alpha-1}(\bm{n}')-b_{\alpha-1}(\bm{n}))}=\bm{n}^{(\alpha-1)}\label{eq: note-on-n-alpha-0}
\end{equation}
Examples of $\bm{n}^{(\alpha,a)}$ are shown in Fig.~\ref{fig: FigS_n_to_alpha-a}. 

When $a\in [0,b_{\alpha-1}(\bm{n}')-b_{\alpha-1}(\bm{n})-1]$, it holds that 
\begin{align}
&b_\alpha(\bm{n}^{(\alpha,a)})
-b_{\alpha-1}(\bm{n}^{(\alpha,a)})\nonumber\\
=&
b_\alpha(\bm{n}^{(\alpha)})
-
b_{\alpha-1}(\bm{n}^{(\alpha)})-a\nonumber\\
=&
b_\alpha(\bm{n}^{(\alpha)})
-
b_{\alpha-1}(\bm{n}')-a\nonumber\\
\ge&
b_\alpha(\bm{n}^{(\alpha)})
-
b_{\alpha-1}(\bm{n})+1\nonumber\\
=&
\underbrace{b_\alpha(\bm{n}^{(\alpha)})
-
b_{\alpha-1}(\bm{n}^{(\alpha)})}_{\ge 1}+1\ge 2.
\end{align}
From Lemma S2, it follows that $\bm{n}^{(\alpha,a)}\sim \bm{n}^{(\alpha,a+1)}$ for 
 $a\in [0,b_{\alpha-1}(\bm{n}')-b_{\alpha-1}(\bm{n})]$ and thus $\bm{n}^{(\alpha)}\sim \bm{n}^{(\alpha-1)}$ (See Eq.~\eqref{eq: note-on-n-alpha-0}) for $\alpha\in [2,2N]$.  
From the relation and Eq.~\eqref{eq: n-alpha-limiting}, $\bm{n}\sim \bm{n}^{(1)}$. Similarly, $\bm{n}'\sim \bm{n}^{'(1)}=\bm{n}^{(1)}$ holds and it follows that $\bm{n}\sim \bm{n}'$. 
\end{proof}
\subsection{Derivation of Eq.~\eqref{eq: UTU}}
\begin{align}
\hat{T}^\dagger\hat{U}\hat{T}&=\hat{U}|_{\hat{P}_j^{(S-f)}\rightarrow \hat{P}_{j-1}^{(S-f)}}\nonumber\\
&=\exp\left[\iu \pi\sum_{j=1}^L\sum_{f=1}^{2S}\sum_{a=0}^{2S-f}jf \hat{P}_{j-1}^{(S-a)}\hat{P}_{j}^{(S-a-f)}\right]\nonumber\\
&=\uwave{\exp\left[\iu \pi\sum_{j=0}^{L-1}\sum_{f=1}^{2S}\sum_{a=0}^{2S-f}jf \hat{P}_{j}^{(S-a)}\hat{P}_{j+1}^{(S-a-f)}\right]}\nonumber\\
&\times\exp\left[\iu \pi\dashuline{\sum_{j=0}^{L-1}\sum_{f=1}^{2S}\sum_{a=0}^{2S-f}f \hat{P}_{j}^{(S-a)}\hat{P}_{j+1}^{(S-a-f)}}\right]\nonumber\\
\end{align}
In the third equality, we replaced the dummy index $j$ by $j+1$. We will show that (i) the Expression underlined by a wavy line is equal to $\hat{U}$ and (ii) that underlined by a dashed line coincides with 
\begin{equation}
\dashuline{\sum_{j=0}^{L-1}\sum_{f=1}^{2S}\sum_{a=0}^{2S-f}f \hat{P}_{j}^{(S-a)}\hat{P}_{j+1}^{(S-a-f)}}=\sum_f f \hat{N}_f.
\label{eq: ii}
\end{equation}
Concerning (i), we find that 
\begin{align}
&\uwave{\exp\left[\iu \pi\sum_{j=0}^{L-1}\sum_{f=1}^{2S}\sum_{a=0}^{2S-f}jf \hat{P}_{j}^{(S-a)}\hat{P}_{j+1}^{(S-a-f)}\right]}\nonumber\\
&=\hat{U}\exp\left[-\iu \pi L\sum_{f=1}^{2S}\sum_{a=0}^{2S-f}f \hat{P}_{L}^{(S-a)}\hat{P}_{L+1}^{(S-a-f)}\right]\nonumber\\
&=\hat{U}.
\end{align}
In the last equality, we have used that  $L$ is even. 

Concerning (ii), we find that the left hand side of Eq.~\eqref{eq: ii} is rewritten as
\begin{align}
&\sum_{j=0}^{L-1}\sum_{1\le a\le a+f\le 2S}f \hat{P}_{j}^{(S-a)}\hat{P}_{j+1}^{(S-a-f)}\nonumber\\
=&\sum_{j=0}^{L-1}\sum_{1\le a\le b\le 2S}(b-a) \hat{P}_{j}^{(S-a)}\hat{P}_{j+1}^{(S-b)}
\label{eq: ii-rewritten}
\end{align}
With use of Eqs.~\eqref{eq: Nf-def-f=2S} and \eqref{eq: Nf-def-f<2S}, the right hand side of Eq.~\eqref{eq: ii} is rewritten as
\begin{align}
&\sum_{f=1}^{2S-1}f\sum_{i=1}^L \left( \sum_{a<f}\hat{P}_{i-1}^{(S-a)}-\sum_{b>f}\hat{P}_{i+1}^{(S-b)}\right)\hat{P}_{i}^{(S-f)}\nonumber\\
&+
2S\sum_{i=1}^L \sum_{a<f}\hat{P}_{i-1}^{(S-a)}\hat{P}_{i}^{(S-f)}\nonumber\\
&=\sum_{i=1}^L\left[\sum_{1\le a<f\le 2S}f  \hat{P}_{i}^{(S-a)}\hat{P}_{i+1}^{(S-f)}
-\sum_{1\le f\le b\le 2S}f \hat{P}_{i}^{(S-f)}\hat{P}_{i+1}^{(S-b)}\right]
\nonumber\\
&=\sum_{i=1}^L\left(\sum_{1\le a<b\le 2S}(b-a)  \hat{P}_{i}^{(S-a)}\hat{P}_{i+1}^{(S-b)}
\right),
\end{align}
which concides with Eq.\eqref{eq: ii-rewritten} and thus the relation Eq.~\eqref{eq: ii} follows. 

\subsection{Derivation of Eq.~\eqref{eq: hprimei}}
We will derive from Eq.~\eqref{eq:  hprimei-first-line} to Eq.~\eqref{eq: hprimei}.
In Eq.~\eqref{eq:  hprimei-first-line}, the operator  $\hat{\mathcal{S}}_i^x (a,b)
\hat{P}_{i-1}^{(S-a)}\hat{P}_{i+1}^{(S-b)}$ does not contain $\hat{P}_{j}^{(S-f)}$ with $j\ne,i,i\pm 1$ and thus the part containing these projection operators in $\hat{U}$ and $\hat{U}^\dagger$ can be dropped. Further $\hat{P}_{i\pm 1}^{(S-f)}$ in $\hat{U}$ and $\hat{U}^\dagger$ can be replaced as
\begin{equation}
\hat{P}_{i-1}^{(S-f)}\rightarrow \hat{1}\delta_{f,a}   
\quad 
\hat{P}_{i+1}^{(S-f)}\rightarrow \hat{1}\delta_{f,b}   
\end{equation}
owing to the presence of $\hat{P}_{i-1}^{(S-a)}\hat{P}_{i+1}^{(S-b)}$
in Eq.~\eqref{eq:  hprimei-first-line}. With use of it, 
the summand in Eq.~\eqref{eq:  hprimei-first-line} reduces to  
\begin{align}
&\hat{U}\hat{\mathcal{S}}_i^x (a,b)
\hat{P}_{i-1}^{(S-a)}\hat{P}_{i+1}^{(S-b)}\hat{U}^\dagger\nonumber\\
=&\hat{u}_i\hat{\mathcal{S}}_i^x (a,b)\hat{u}_i^\dagger
\hat{P}_{i-1}^{(S-a)}\hat{P}_{i+1}^{(S-b)}\label{eq: ui-S-uid-PP}
\end{align}
with 
\begin{equation}
\hat{u}_i=\exp\left[\iu \pi \sum_{c=a}^b ((c-a)(i-1)+(b-c)i)\hat{P}_i^{(S-c)}\right].     
\end{equation}
When $a\ge b$, $\hat{u}_i=\hat{1}$ and thus 
\begin{align}
\hat{u}_i\hat{\mathcal{S}}_i^x (a,b)\hat{u}_i^\dagger
=\hat{\mathcal{S}}_i^x (a,b) \label{eq: ui-Sx-ui-a>b}
\end{align}
for $a\ge b$. 
We thus consider the case when $a<b$. The operator  
\begin{equation}
\hat{u}_i\hat{\mathcal{S}}_i^x (a,b)\hat{u}_i^\dagger
\label{eq: single-site-operator}
\end{equation}
is a single site operator for the $i$-th site. From the operator contents in $\hat{\mathcal{S}}_i^x (a,b)$, it follows that 
\begin{equation}
\langle n_i|\hat{u}_i\hat{\mathcal{S}}_i^x (a,b)\hat{u}_i^\dagger|n'_i\rangle=0    
\end{equation}
unless $n_i=n'_i\pm 1$. With use of this property, Eq.~\eqref{eq: single-site-operator} is rewritten as
\begin{align}
&\hat{u}_i\hat{\mathcal{S}}_i^x (a,b)\hat{u}_i^\dagger
\nonumber\\
&=\sum_{0\le k,k+1\le 2S}\hat{P}_i^{(S-k)}\hat{u}_i\hat{\mathcal{S}}_i^x (a,b)\hat{u}_i^\dagger\hat{P}_i^{(S-k-1)}\label{eq: single-site-operator_k+1}\\
&+
\sum_{0\le k,k-1\le 2S}\hat{P}_i^{(S-k)}\hat{u}_i\hat{\mathcal{S}}_i^x (a,b)\hat{u}_i^\dagger\hat{P}_i^{(S-k+1)}. \label{eq: single-site-operator_k-1}
\end{align}
In Eq.~\eqref{eq: single-site-operator_k+1}, the operator $\hat{P}_i^{(S-c)}$ in $\hat{u}_i$ can be replaced by $
\hat{1}\delta_{k,c}$ owing to the presence of $\hat{P}_i^{(S-k)}$. With use of it, $\hat{u}_i$ 
in Eq.~\eqref{eq: single-site-operator_k+1} can be reduced to
\begin{equation}
\hat{u}_i\rightarrow 
(-1)^{bi-a(i-1)-k}\hat{1}.\label{eq: ui-reduction}
\end{equation}
Similarly, 
$\hat{P}_i^{(S-c)}$ in $\hat{u}_i^\dagger$ can be replaced by
$\hat{1}\delta_{k+1,c}$ and $\hat{u}_i^\dagger$ in Eq.~\eqref{eq: single-site-operator_k+1} reduces to  
\begin{equation}
\hat{u}_i^\dagger\rightarrow 
(-1)^{-bi+a(i-1)+k+1}\hat{1}\label{eq: uid-reduction}.
\end{equation}
From \eqref{eq: ui-reduction} and \eqref{eq: uid-reduction}, 
Eq.~\eqref{eq: single-site-operator_k+1} reduces to 
\begin{equation}
-\sum_{0\le k,k+1\le 2S}\hat{P}_i^{(S-k)}\hat{\mathcal{S}}_i^x (a,b)\hat{P}_i^{(S-k-1)}
\label{eq: single-site-operator_k+1-reduction}
\end{equation}
Similarly, 
Eq.~\eqref{eq: single-site-operator_k-1} becomes
\begin{equation}
-\sum_{0\le k,k-1\le 2S}\hat{P}_i^{(S-k)}\hat{\mathcal{S}}_i^x (a,b)\hat{P}_i^{(S-k+1)}.
\label{eq: single-site-operator_k-1-reduction}
\end{equation}
From \eqref{eq: single-site-operator_k+1-reduction} and \eqref{eq: single-site-operator_k-1-reduction}, 
\begin{align}
&\hat{u}_i\hat{\mathcal{S}}_i^x (a,b)\hat{u}_i^\dagger
\nonumber\\
=&-\sum_{0\le k,k+1\le 2S}\hat{P}_i^{(S-k)}\hat{\mathcal{S}}_i^x (a,b)\hat{P}_i^{(S-k-1)}\nonumber\\
&-\sum_{0\le k,k-1\le 2S}\hat{P}_i^{(S-k)}\hat{\mathcal{S}}_i^x (a,b)\hat{P}_i^{(S-k+1)}\nonumber\\
&=-\hat{\mathcal{S}}_i^x (a,b). \label{eq: ui-Sx-ui-a<b}
\end{align}
for $a<b$. From Eqs.~\eqref{eq:  hprimei-first-line}, \eqref{eq: ui-S-uid-PP}, \eqref{eq: ui-Sx-ui-a<b}, and \eqref{eq: ui-Sx-ui-a>b}, Eq.~\eqref{eq: hprimei} follows.

\subsection{Proof of Lemma 6}
As a statement equivalent to Eq.~\eqref{eq: statement-of-Lemma6}, we prove 
\begin{equation}
\langle T(\bm{n})|(-\hat{\mathcal{H}}'_{\rm p0})^l|\bm{n}\rangle\ne 0\label{eq: statement-of-Lemma6-rewritten}
\end{equation}
because the off-diagonal matrix elements of $\hat{\mathcal{H}}_{{\rm p}}$ coincide with those of $\hat{\mathcal{H}}'_{{\rm p}}$ up to 
an overall sign and the latter stems from those of $\hat{\mathcal{H}}'_{\rm p0}$. 
Further, all the off-diagonal matrix elements of $-\hat{\mathcal{H}}'_{\rm p0}$ are non-negative and thus it suffices to find a set of the intermediate states $\{\bm{n}^{(1)},\bm{n}^{(2)},\cdots,\bm{n}^{(l-1)}\}$
satisfying  
\begin{equation}
\langle \bm{n}^{(a)}|\hat{\mathcal{H}}'_{\rm p0}|\bm{n}^{(a-1)}\rangle\ne 0,\quad\mbox{for }a=[1,l],    
\end{equation}
where $\bm{n}^{(0)}$ and $\bm{n}^{(l)}$, respectively, read as 
$\bm{n}$ and $T(\bm{n})$. 

We find, from Eqs.~\eqref{eq: hprimei} and \eqref{eq: mathcalSix}, the support of the local Hamiltonian $\hat{h}'_i$. $\hat{h}'_i|\bm{n}\rangle$ can be nonzero only when the following two conditions are satisfied: 
\begin{subequations}
\begin{align}
&n_{i-1}\ne n_{i+1}\label{eq: condition-1}\\    
&{\rm min}(n_{i-1},n_{i+1})\le n_i\le {\rm max}(n_{i-1},n_{i+1}).\label{eq: condition-2}  
\end{align}
\end{subequations}
Let $\mathcal{V}(i)$ be the set of $\bm{n}$ satisfying Eqs.~\eqref{eq: condition-1} and \eqref{eq: condition-2}. The condition Eq.~\eqref{eq: condition-1} comes from the factor $|a-b|\hat{P}_{i-1}^{(S-a)}\hat{P}_{i+1}^{(S-b)}$ in Eq.~\eqref{eq: hprimei}
and the other condition Eq.~\eqref{eq: condition-2} does from the factor $\hat{P}_{i-1}^{(S-a)}\hat{P}_{i+1}^{(S-b)}\left(\sum_{k={\rm min}(a,b)}^{{\rm max}(a,b)}\hat{P}_i ^{(S-a)}\right)$ in Eq.~\eqref{eq: hprimei} with Eq.~\eqref{eq: mathcalSix}. 

The local Hamiltonian $\hat{h}'_i$ commutes with $\hat{S}_j^z$ for $j\ne i$ and thus $\langle\bm{n}'|\hat{h}'_i |\bm{n}\rangle$ reduces to
\begin{align}
&\langle\bm{n}'| \hat{h}'_i |\bm{n}\rangle\nonumber\\
&=|n_{i-1}-n_{i+1}|\langle n_i'|\hat{S}_i^x|n_i\rangle_i \prod_{j(\ne i)}\delta_{n'_j,n_j},&\mbox{for }\bm{n}', \bm{n}\in \mathcal{V}(i), \label{eq: me-of-hi-to-a}
\end{align}
which can be summarized as 
\begin{align}
&\langle\bm{n}'| \hat{h}'_i|\bm{n}\rangle\nonumber\\
&\left\{
\begin{array}{cc}
>0,&\bm{n}',\bm{n}\in \mathcal{V}(i) \mbox{ and } n'_j=n_j\pm \delta_{i,j},\mbox{ for }j\in [1,L]\\
=0,&\mbox{otherwise.}
\end{array}
\right.
\label{eq: menprimen}    
\end{align}
From Eq.~\eqref{eq: menprimen}, it follows that 
\begin{align}
&\langle\bm{n}'| \hat{h}'_k|\bm{n}\rangle=0,\mbox{ for }k\ne i  \\  
&\langle\bm{n}'|\hat{\mathcal{H}}'_{\rm p0}|\bm{n}\rangle=-D
\langle\bm{n}'| \hat{h}'_i|\bm{n}\rangle<0,\nonumber\\
&\mbox{when }\bm{n}',\bm{n}\in \mathcal{V}(i) \mbox{ and } n'_j=n_j\pm \delta_{i,j},\mbox{ for }j\in [1,L].\label{eq: me-of-HM0}
\end{align}
Below we provide a proof of Lemma 6 using Eq.~\eqref{eq: me-of-hi-to-a}. 
It is convenient for this purpose to parametrize $\bm{n}$ in the following way. 
For a given $\bm{n}$, let $B=B(\bm{n})$ be the set of $i$ such that $n_{i-1}\ne n_i$ and $d_B$ be the number of elements of $B$. We label 
these elements in $B(\bm{n})=\{b_\alpha\}$ 
in increasing order
\begin{equation}
1\le b_1<b_2\cdots<b_{d_B}\le L.
\end{equation}
We define $m_\alpha\in [0,2S]$ for $\alpha=[1,d_B]$ by 
\begin{equation}
m_\alpha=n_i,\mbox{ with }i=b_\alpha.    
\end{equation}
We also define $b_{d_B +1}$ and $m_{d_B +1}$, respectively as, $b_1 +L$ and $m_{1}$. 
By definition, $m_\alpha\ne m_{\alpha+1}$ for  $\alpha=[1,d_B]$.   
When $L=8$ and $\bm{n}=00133000$, for example (see Fig.~\ref{fig:  example-balpha-malpha-and-nprime-higherS}), $d_B=3$ and
\begin{subequations}
\begin{align}
&b_1=3 ,\quad b_2=4,\quad b_3=6,\quad b_4=11 \\    
&m_1=1 ,\quad m_2=3,\quad m_3=0,\quad m_4=1.     
\end{align}
\label{eq: example-balpha-malpha-higherS}
\end{subequations}
\begin{center}
\begin{figure}
\includegraphics[width=0.25\columnwidth,pagebox=artbox]{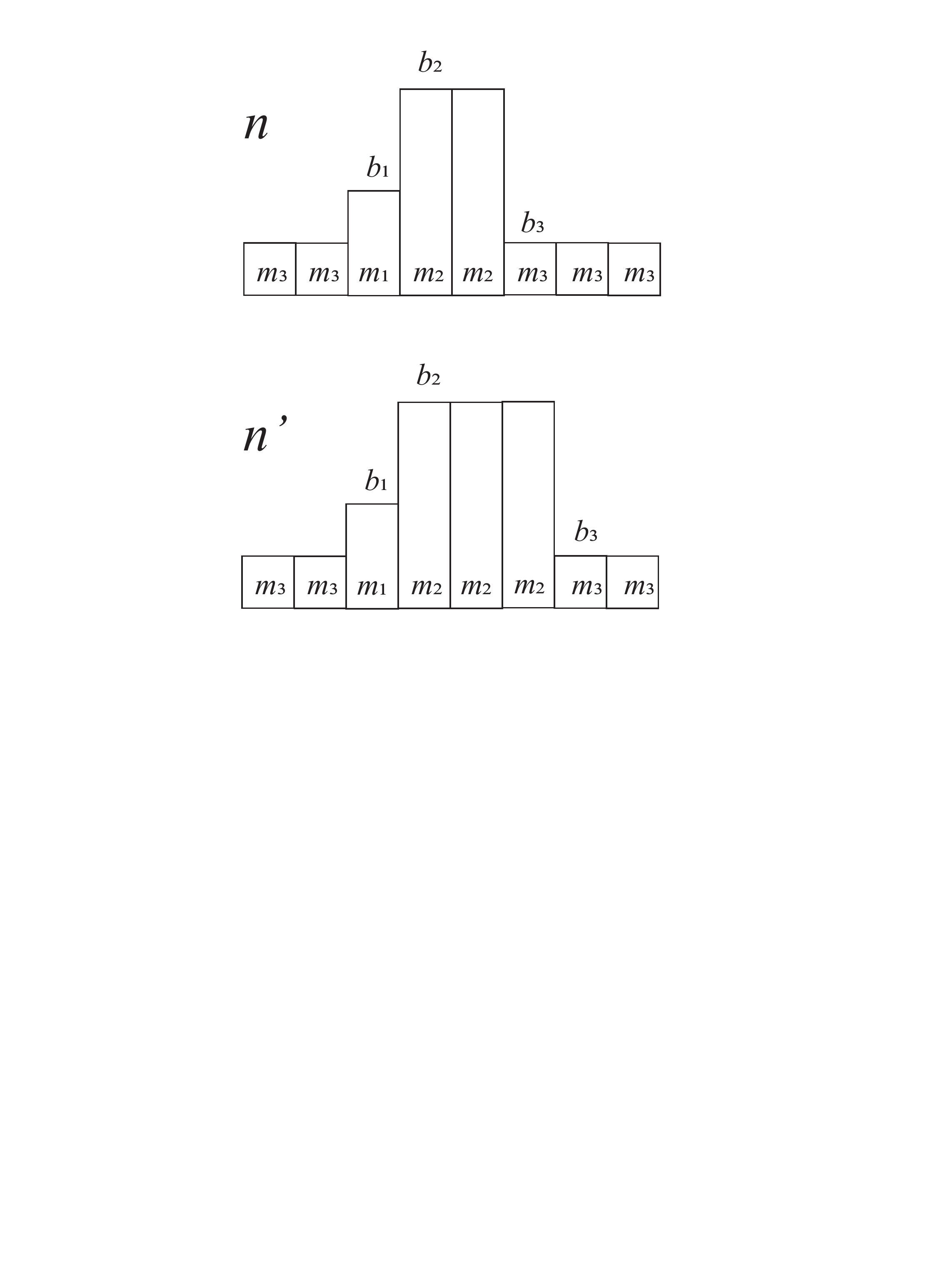}
\caption{Upper panel: Example Eq.~\eqref{eq: example-balpha-malpha-higherS} of $\{b_\alpha,m_\alpha\}_{\alpha=1}^{d_B}$ for $\bm{n}=00133000$. $d_B=3$ and $L=8$. Lower panel: Example of $\bm{n}'$ defined by Eq.~\eqref{eq: n-prime-higherS}.}
\label{fig: example-balpha-malpha-and-nprime-higherS}
\end{figure}
\end{center}
In 
these notations, the relation between $\bm{n}$ and $T(\bm{n})$
reads
\begin{subequations}
\begin{align}
b_\beta(T(\bm{n}))&=b_\beta(\bm{n})+1\\
m_\beta(T(\bm{n}))&=m_\beta(\bm{n})
\end{align}
\label{eq: b-shift-n-and-Tn}
\end{subequations}
for $\beta\in [1,d_B]$.     

To prove 
Lemma 6, we consider the following two cases separately: 
\begin{itemize}
    \item Case 1. $^\exists\alpha\in [1,d_B],\quad b_{\alpha+1}-b_{\alpha}\ge 2$.
    \item Case 2. $^\forall\alpha\in [1,d_B],\quad  b_{\alpha+1}-b_{\alpha}=1 $ (mod $L$).

\end{itemize}
{\bf Case 1}\\
For 
case 1, we consider 
the situation where $^\exists\alpha\in [1,d_B-1],\quad b_{\alpha+1}-b_{\alpha}\ge 2$. For this case, 
the portion of the spin configuration $\bm{n}$ concerning the three consecutive sites $i=b_\alpha-1,b_\alpha,b_\alpha+1$ is %
\begin{align}
\bm{n}=(\cdots,m_{\alpha-1},m_\alpha,m_\alpha,\cdots)
\label{eq: n-three-sites}
\end{align}
from which we see that $\bm{n}\in \mathcal{V}(b_\alpha)$ because Eq.~\eqref{eq: n-three-sites} satisfies Eqs.~\eqref{eq: condition-1} and \eqref{eq: condition-2} with $i=b_\alpha$. We assume that $m_{\alpha-1}-m_\alpha>0$. The case where  $m_{\alpha-1}<m_\alpha$ can be dealt with in a way similar to the following argument. 
We show that the matrix element $ \langle \bm{n}'|(-\hat{\mathcal{H}}_{\rm p0})^{(m_{\alpha-1}-m_\alpha)}|\bm{n}\rangle> 0$, 
where 
\begin{equation}
n'_j=n_j +(m_{\alpha-1}-m_\alpha)\delta_{i,j}.  
\label{eq: n-prime-higherS}
\end{equation}
In a way similar to Eq.~\eqref{eq: n-three-sites}$, \bm{n}'$ is expressed as
\begin{align}
\bm{n}'=(\cdots,m_{\alpha-1},m_{\alpha-1},m_\alpha,\cdots)
\label{eq: nprime-three-sites}
\end{align}
and 
\begin{subequations}
\begin{align}
b_\beta(\bm{n}')&=b_\beta(\bm{n})+\delta_{\beta,\alpha}\\
m_\beta(\bm{n}')&=m_\beta(\bm{n})
\end{align}
\label{eq: b-nprime-n-relation}
\end{subequations}
for $\beta\in [1,d_B]$.     
Example of $\bm{n}'$ is shown in the lower panel in 

We introduce a series of states $\{\bm{n}^{(0)},\bm{n}^{(1)},\bm{n}^{(2)},\cdots,\bm{n}^{(m_{\alpha-1}-m_\alpha)}\}$ as
\begin{equation}
n^{(a)}_j=n_j +a\delta_{i,j},\quad\mbox{for }j\in [1,L], \label{eq: nja-def}   
\end{equation}
i.e.
\begin{align}
\bm{n}^{(a)}=(\cdots,m_{\alpha-1},m_\alpha+a,m_\alpha,\cdots)
\label{eq: n(a)-three-sites}
\end{align}
for $a=[0,m_{\alpha-1}-m_\alpha]$.
Note that $\bm{n}^{(0)}=\bm{n}$ and $\bm{n}^{(m_{\alpha-1}-m_\alpha)}=\bm{n}'$.  
We see that $\bm{n}^{(a)}\in \mathcal{V}(b_\alpha)$ for $a\in [0,m_{\alpha-1}-m_\alpha]$ and 
\begin{equation}
n^{(a)}_j=n^{(a-1)}_j+\delta_{i,j},\quad\mbox{for }j\in [1,L].    
\end{equation}
\begin{center}
\begin{figure}
\includegraphics[width=0.25\columnwidth,pagebox=artbox]{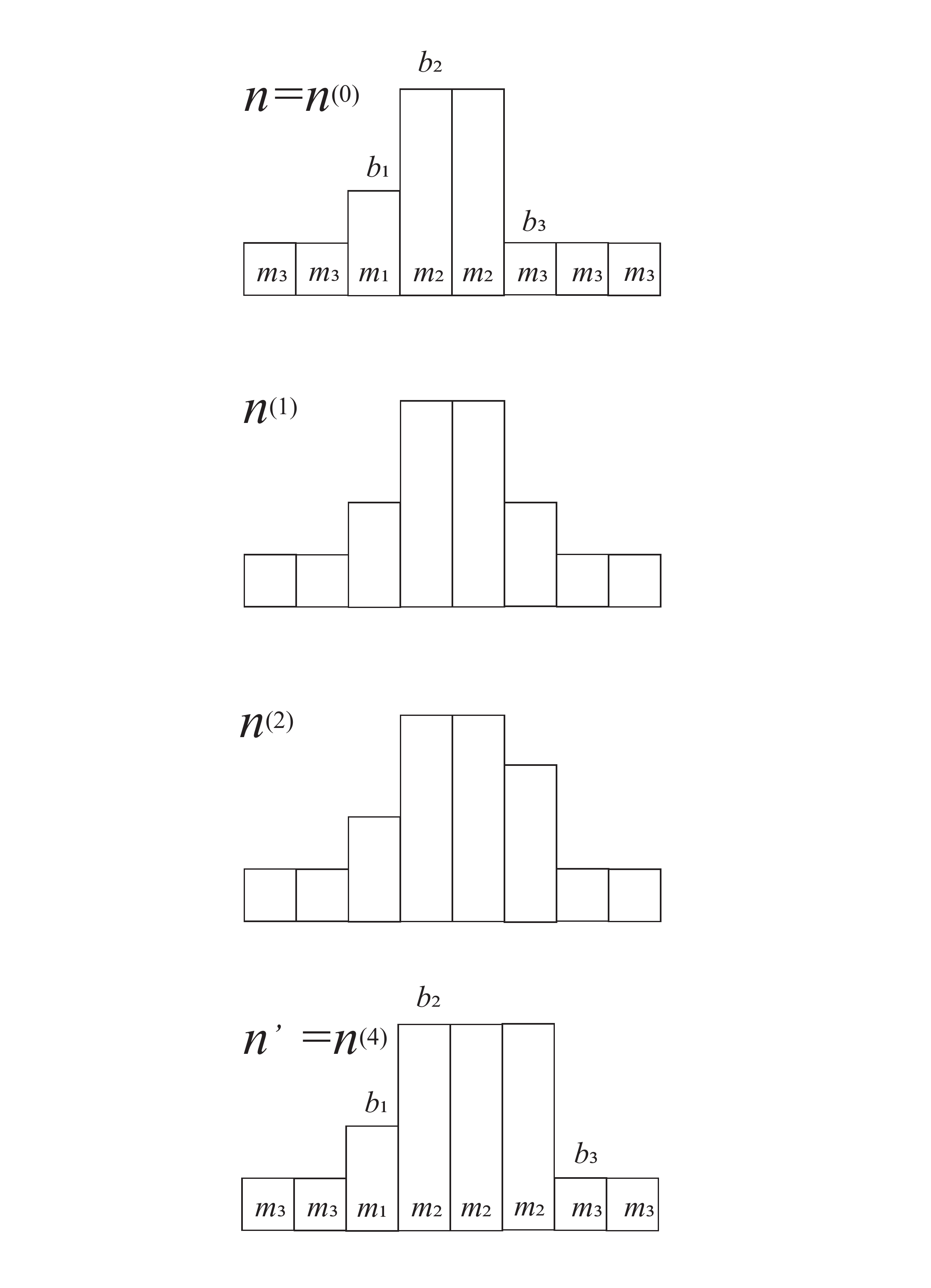}
\caption{Examples of $n^{(a)}$ defined by Eq.~\eqref{eq: nja-def} or \eqref{eq: n(a)-three-sites} where $\bm{n}=00133000$, $\bm{n}'=00133300$, $i=6$, and $L=8$. }
\label{fig: FigS_n-a-higherS}
\end{figure}
\end{center}
From this observation and Eq.~\eqref{eq: me-of-HM0}, we find that 
\begin{equation}
\langle\bm{n}^{(a)}|\hat{\mathcal{H}}'_{\rm p0}|\bm{n}^{(a-1)}\rangle<0,\quad\mbox{for }a\in [1,m_{\alpha-1}-m_\alpha],  
\end{equation}
from which 
\begin{equation}
\langle\bm{n}'|(-\hat{\mathcal{H}}'_{\rm p0})^{m_{\alpha-1}-m_\alpha}|\bm{n}\rangle
>0 \label{eq: HM0-nprime-and-n} 
\end{equation}
follows for $\bm{n}$ (Eq.~\eqref{eq: n-three-sites}) and $\bm{n}'$ (Eq.~\eqref{eq: nprime-three-sites}) .
From Eqs.~\eqref{eq: b-nprime-n-relation} and \eqref{eq: HM0-nprime-and-n}, we see that multiple action{\color{black}s} of $\hat{\mathcal{H}}'_{\rm p0}$ on $|\bm{n}\rangle$ can shift one of the boundaries $b_\alpha(\bm{n})$ by 
one site to the right when $b_{\alpha+1}(\bm{n})-b_\alpha(\bm{n})\ge 2$. 
We note that $b_{\alpha}(\bm{n}')-b_{\alpha-1}(\bm{n}')=b_{\alpha}(\bm{n})-b_{\alpha-1}(\bm{n})+1\ge 2$ and thus find that we can shift $b_{\alpha-1}(\bm{n}')$ by 
one site to the right by 
multiple action{\color{black}s} of 
$\hat{\mathcal{H}}'_{\rm p0}$ on $|\bm{n}'\rangle$, viz, 
\begin{equation}
\langle\bm{n}''|(-\hat{\mathcal{H}}'_{\rm p0})^{m_{\alpha-2}-m_{\alpha-1}}|\bm{n}'\rangle
>0 \label{eq: HM0-ndoubleprime-and-nprime} 
\end{equation}
for $\bm{n}'$ (Eq.~\eqref{eq: nprime-three-sites}) and $\bm{n}''$ satisfying
\begin{subequations}
\begin{align}
b_\beta(\bm{n}'')&=b_\beta(\bm{n}')+\delta_{\beta,\alpha-1}\\
m_\beta(\bm{n}'')&=m_\beta(\bm{n}')
\end{align}
\label{eq: b-nprime-n-relation-case1}
\end{subequations}
for $\beta\in [1,d_B]$. Note that $b_{\alpha-1}(\bm{n}'')-b_{\alpha-2}(\bm{n}'')=b_{\alpha-1}(\bm{n}')-b_{\alpha-2}(\bm{n}')+1\ge 2$ and thus we can shift $b_{\alpha-2}(\bm{n}'')$ by 
{\color{black}one site} to the right in a way similar to the above procedure. By repeating these procedures, we can shift all $b_\beta(\bm{n})$ for 
$\beta\in [1,d_B]$ by 
{\color{black}one site} to the right and arrive at $T(\bm{n})$ satisfying Eq.~\eqref{eq: b-shift-n-and-Tn}. \\
\ \\
{\bf Case 2}\\
We discuss 
Case 2, where $\beta\in [1,d_B],\quad  b_{\beta+1}-b_{\beta}\equiv 1 $ (mod $L$). In 
Lemma 6, we consider $\bm{n}\in V_\mu(\{N_f\})$ with $d(V_\mu(\{N_f\}))>1$ and thus there exists $\alpha\in [1,d_B]$ such that $\bm{n}\in \mathcal{V}(b_\alpha)$. In this case, {\color{black}the} spin configuration in $\bm{n}$ for consecutive three sites $i=b_\alpha-1,b_\alpha,b_\alpha+1$ is given by
\begin{align}
\bm{n}=(\cdots,m_{\alpha-1},m_\alpha,m_{\alpha+1},\cdots).
\label{eq: n-three-sites-case2}
\end{align}
We assume that $m_{\alpha-1}>m_\alpha>m_{\alpha+1}$. The case where $m_{\alpha-1}<m_\alpha<m_{\alpha+1}$ can be discussed in a similar way. 

\begin{center}
\begin{figure}
\includegraphics[width=0.5\columnwidth,pagebox=artbox]{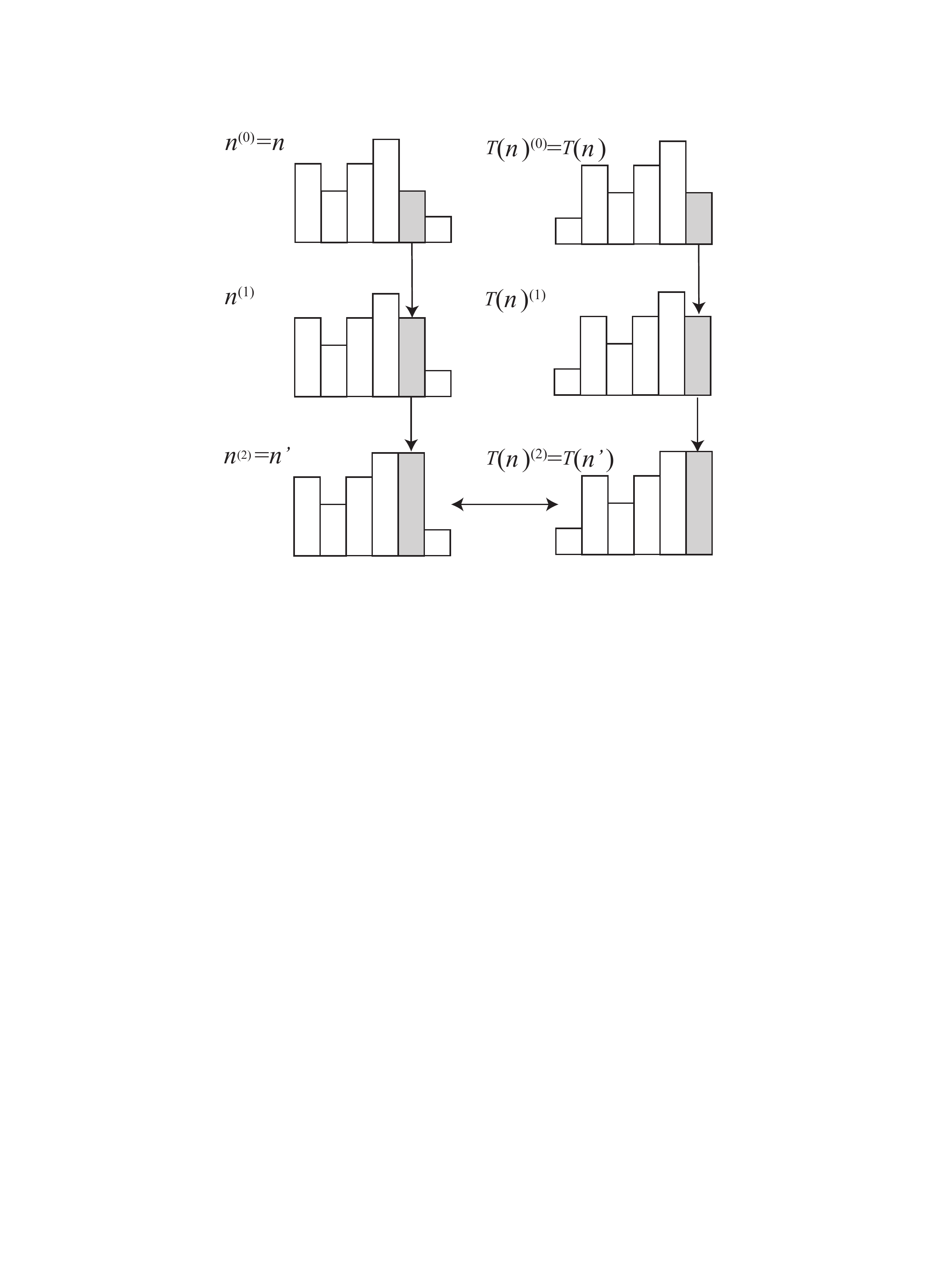}
\caption{Examples of Eqs.~\eqref{eq: HM0-nprime-and-n-case2} (left column), Eq.~\eqref{eq: statement-of-Lemma6-rewritten-nprime} (right column), and Eq.~\eqref{eq: me-Tn-Tnprime} (two figures in the bottom) for Case 2.}
\label{fig: FigS_lemma6-case2}
\end{figure}
\end{center}

We introduce a series of the states $\{\bm{n}^{(1)},\bm{n}^{(2)},\cdots,\bm{n}^{(m_{\alpha-1}-m_\alpha)}\}$ satisfying Eq.~\eqref{eq: nja-def}. In the present case, Eq.~\eqref{eq: n(a)-three-sites} should be replaced by
\begin{align}
\bm{n}^{(a)}=(\cdots,m_{\alpha-1},m_\alpha+a,m_{\alpha+1},\cdots)
\label{eq: n(a)-three-sites-case2}
\end{align}
for $a=[1,m_{\alpha-1}-m_\alpha]$. We can show that 
\begin{equation}
\langle\bm{n}'|(-\hat{\mathcal{H}}'_{\rm p0})^{m_{\alpha-1}-m_\alpha}|\bm{n}\rangle
>0 \label{eq: HM0-nprime-and-n-case2} 
\end{equation}
for $\bm{n}'=\bm{n}^{(m_{\alpha-1}-m_\alpha)}$ in a way similar to the proof of Eq.~\eqref{eq: HM0-nprime-and-n}. Spin configuration in $\bm{n}'$ for consecutive three sites $i=b_\alpha-1,b_\alpha,b_\alpha+1$ is given by
\begin{align}
\bm{n}'=(\cdots,m_{\alpha-1},m_{\alpha-1},m_{\alpha+1},\cdots).
\label{eq: n-prime-three-sites-case2}
\end{align}
and thus the argument for Case 1 is applicable to $\bm{n}'$, i.e., there exists a positive integer $l$ such that 
\begin{equation}
\langle T(\bm{n}')|(-\hat{\mathcal{H}}'_{\rm p0})^l|\bm{n}'\rangle>0\label{eq: statement-of-Lemma6-rewritten-nprime}.
\end{equation}
Further 
\begin{align}
&\langle T(\bm{n})|(-\hat{\mathcal{H}}'_{\rm p0})^{m_{\alpha-1}-m_\alpha}|T(\bm{n}')\rangle\nonumber\\
&=\langle \bm{n}|(-\hat{\mathcal{H}}'_{\rm p0})^{{m_{\alpha-1}-m_\alpha}}|\bm{n}'\rangle>0
\label{eq: me-Tn-Tnprime}
\end{align}
follows from translational invariance of $\hat{\mathcal{H}}'_{\rm p0}$ and Eq.~\eqref{eq: HM0-nprime-and-n-case2}.  Combining Eqs.~\eqref{eq: HM0-nprime-and-n-case2}, \eqref{eq: statement-of-Lemma6-rewritten-nprime}, and \eqref
{eq: me-Tn-Tnprime}, we arrive at Lemma 6 for Case 2. 
Figure~\ref{fig: FigS_lemma6-case2} schematically shows that 
Eqs.~\eqref{eq: HM0-nprime-and-n-case2} (left column), \eqref{eq: statement-of-Lemma6-rewritten-nprime} (right column), and \eqref{eq: me-Tn-Tnprime} (two figures in the bottom).
\end{document}